\documentclass[letterpaper,twocolumn,10pt]{article}
\usepackage{usenix-2020-09}
\usepackage{hyperref}
\usepackage{tikz}
\usepackage{amsmath}
\usepackage{xurl}
\usepackage{listings}
\usepackage{subcaption}
\usepackage{mdframed}
\usepackage{nopageno}

\newcommand{\etal}{\textit{et.al}}

\begin{document}


\title{\Large \bf Large Language Models for Code Analysis: Do LLMs Really Do Their Job?}

\author{
{\rm Chongzhou Fang$^1$, Ning Miao$^1$, Shaurya Srivastav$^1$, Jialin Liu$^2$, Ruoyu Zhang$^1$, Ruijie Fang$^1$,} \and
{\rm Asmita$^1$, Ryan Tsang$^1$, Najmeh Nazari$^1$, Han Wang$^2$, and Houman Homayoun$^1$} \\ \\
$^1$University of California, Davis
\quad $^2$Temple University
} 

\maketitle

\begin{abstract}

Large language models (LLMs) have demonstrated significant potential in the realm of natural language understanding and programming code processing tasks. Their capacity to comprehend and generate human-like code has spurred research into harnessing LLMs for code analysis purposes. However, the existing body of literature falls short in delivering a systematic evaluation and assessment of LLMs' effectiveness in code analysis, particularly in the context of obfuscated code.

This paper seeks to bridge this gap by offering a comprehensive evaluation of LLMs' capabilities in performing code analysis tasks. Additionally, it presents real-world case studies that employ LLMs for code analysis. Our findings indicate that LLMs can indeed serve as valuable tools for automating code analysis, albeit with certain limitations. Through meticulous exploration, this research contributes to a deeper understanding of the potential and constraints associated with utilizing LLMs in code analysis, paving the way for enhanced applications in this critical domain.
\end{abstract}

\section{Introduction}\label{SecIntro}
The evolution of machine learning has brought major changes to a variety of fields in the past few decades~\cite{he2016deep,vaswani2017attention, topol2019high,fang2022towards}.
In recent years, the emergence of large language models (LLMs) has revolutionized the field of natural language processing and brought great changes to many other areas. Ever since the access of ChatGPT~\cite{chatgpt} becomes publicly available in late 2022, there is a rocketing number of users as well as publicly-available services utilizing LLM APIs provided by OpenAI. These models have demonstrated their abilities to understand and process inputs including but not limited to natural languages, and have been involved in different types of tasks. A typical use case of LLMs is text generation, since it has the ability to produce high-quality natural language outputs~\cite{zhao2023survey}. This has caused a heated debate in academia and eventually resulted in major publishers announcing new regulations regarding the usage of generative AI~\cite{acm, ieee}. Besides, LLMs have also manifested their extraordinary potential in tasks related to programming, including code generation~\cite{vaithilingam2022expectation} and code comprehension~\cite{yuan2023evaluating}. LLM-based assistant tools like Github Copilot~\cite{copilot} have received tremendous attention and have been widely employed by programmers all around the world. OpenAI has recently enabled more features in ChatGPT~\cite{seehearspeak}, making LLMs more versatile and capable of engaging in more real-world tasks.


Amidst all possibilities of LLMs, code analysis and obfuscation are tasks that are receiving more and more attention from computer scientists. Code analysis~\cite{binkley2007source} is an important task in software engineering, including code quality assessment~\cite{salman2015students}, vulnerabilities detection~\cite{xie2006static}, etc. The huge size and complex architecture of modern software systems motivate the wide usage of automated code analysis, especially in security-related scenarios such as software reverse engineering~\cite{canfora2011achievements}. In response, malicious parties have leveraged code obfuscation techniques to obscure source code and avoid being detected. In response, researchers have proposed automated code analysis tools like Abstract Syntax Tree (AST)~\cite{neamtiu2005understanding} to analyze both non-obfuscated and obfuscated code. However, such tools are highly dependent on the pre-defined data types and not suitable for general software code analysis. Due to the great potential of LLMs in these areas, researchers have started using LLMs to perform related tasks~\cite{bubeck2023sparks,yuan2023evaluating}. However, as pointed out by a recent survey~\cite{barrett2023identifying}, there are few systematic analytical studies on this topic.

In this paper, we focus on evaluating the ability of LLMs to analyze input code samples, and test if LLMs can be utilized for defensive analysis tasks. We first construct a dataset containing code from real-world programs (on Github, etc.) and their obfuscated versions to feed to LLMs, then we analyze the outcomes and present our obtained key findings. We aim to answer two critical research questions through a systematic analysis:
\begin{description}
    \item[RQ1:] Do LLMs understand source code?
    \item[RQ2:] Can LLMs comprehend obfuscated code or code with low readability?
\end{description}
After that, as real-world case studies, we showcase if publicly-available LLMs (e.g. ChatGPT) can be utilized for security-related tasks targeting real-world malicious software.

The contributions of this paper are listed as follows:
\begin{enumerate}
    \item We construct two datasets consisting of code samples from popular programming languages and obfuscated samples from these languages for our evaluation;
    \item We systematically evaluate the performance of publicly available state-of-the-art LLMs, including the most popularly used GPT and LLaMA families on our dataset, and present our findings;
    \item We conduct case studies using LLMs to analyze real-world malicious software to showcase the capability and limitations of LLMs in these tasks.
\end{enumerate}
To the best of our knowledge, this is the first paper that provides a systematic evaluation of the ability of LLMs to comprehend code and code with obfuscation.

The remainder of this paper is organized as follows. We first introduce related background knowledge in Section~\ref{SecBackground}. Then we elaborate our experiment settings of our analysis in Section~\ref{SecEvalSet}. Essential results and findings are presented in Section~\ref{SecResult}. In Section~\ref{SecCaseStudy}, we perform case studies and show how LLMs can be utilized to assist code analysis in practical cases. Then we briefly discuss future works in Section~\ref{SecDiss}. Related works and a conclusion are provided in Section~\ref{SecRelWork} and Section~\ref{SecConc}, respectively. Our online appendix is available at: \url{https://github.com/aseec-lab/llms-for-code-analysis}.

\section{Background}\label{SecBackground}
In this part, we will provide an overview of the technical background of large language models, code analysis, and code obfuscation.
\subsection{Large Language Models}
Large language models (LLMs) are a groundbreaking innovation in the field of artificial intelligence, representing a significant milestone in natural language understanding and generation. Most LLMs are built on deep learning architectures, particularly transformer architectures and self-attention mechanisms~\cite{vaswani2017attention}, which are trained on massive datasets containing a diverse range of text from the internet, books, and Github. This extensive training enables them to grasp the intricacies of languages, including grammar, context, and semantics. Both companies and researchers have launched LLMs-driven products and pre-trained models, such as OpenAI GPT series~\cite{openai2023gpt4,solaiman2019release}, Meta LLaMA~\cite{touvron2023llama}, Standford Alpaca~\cite{taori2023alpaca}, etc. One of the critical features of the LLMs is their ability to perform a wide range of natural language understanding tasks, including text summarization, translation, sentiment analysis, question-answering, etc. Besides natural language understanding tasks, researchers also find that LLM models have significant performance across various domains, from security and code generation~\cite{chen2021evaluating} to healthcare~\cite{zhang2023privee,hosseini2022convolution,cascella2023evaluating} and education~\cite{kasneci2023chatgpt}. These LLM-driven applications have the potential to automate analysis tasks and provide instant information and assistance~\cite{araci2019finbert}. Moreover, LLMs are being shown to have direct utility for the purposes of code explanation and summarization, which have immediate applications in education and industry. Leinonen et al. \cite{Leinonen2023-ComparingCodeExplanations} have explored the use of LLMs in generating code explanations for the purposes of computer science (CS) education, finding that LLM-generated explanations are viewed as more accurate and comprehensible than student-written explanations. Ahmed et al. \cite{Ahmed2023-FewshotTrainingLLMs} leverage project-specific few-shot training to improve code summarization, demonstrating improvements across a variety of languages and projects. This approach significantly reduces the time programmers need to familiarize themselves with a codebase.



\subsection{Code Analysis}
Code analysis is a process in software engineering to examine source code, byte code, or binary code to ensure quality, reliability, and security. Code analysis has become a crucial practice to assist developers in identifying and addressing problems early in the software development life cycle. The increasingly large scale of modern software challenges motivates researchers to propose automated tools. Prior works have proposed to extract structural relationships, including inheritance, association, friend relationships, interface hierarchies, attributes, data types, etc., from source code~\cite{neamtiu2005understanding} to build Abstract Syntax
Tree (AST) for pattern-matching~\cite{zhang2019novel} to detect vulnerabilities~\cite{cao2021bgnn4vd} and malicious activities~\cite{livshits2010zozzle}. However, the approach is highly dependent on the pre-defined data types and can not be used for general code analysis. In response, some researchers propose to extract features from source code and leverage machine learning to build a vulnerability detection model \cite{zhong2017seq2sql}. 

\subsection{Code Generation}
Code generation has been one of the most popular applications of LLMs since the introduction of GPT-3 OpenAI's introduction of the Codex model \cite{chen2021evaluating}. Subsequent advances in this area have had major implications on how programming is now taught, practiced, and evaluated. Prather \textit{et al.} \cite{Prather2023-RobotsAreHere} noted in their meta-analysis of LLM use in CS education that LLMs perform similar if not better than the average student on code writing tasks, and have since sought to incorporate such tools into CS curriculum \cite{Prather2024-InteractionsPromptProblems} in anticipation of more widespread adoption. A number of studies have also been conducted on practical LLM-based code completion tools like Github's Copilot \cite{copilot}, characterizing common interaction models \cite{Barke2023-GroundedCopilotHow} and experiences \cite{Bird2023-TakingFlightCopilot}. Such tools continue to be enhanced as well, with improvements made in code completion for repository-level projects \cite{Bairi2023-CodePlanRepositorylevelCoding, Li2024-EnhancingLLMBasedCoding} and automated program repair \cite{Wei2023-CopilotingCopilotsFusing, Xia2023-ConversationalAutomatedProgram}, expanding the scope of LLM-based generation tools. Notably, Bird et al. \cite{Bird2023-TakingFlightCopilot} have observed in their investigation that while Copilot can improve productivity and creativity for users, the tool can also be somewhat detrimental to security and programmer understanding, which is part of the motivation for our study.

\subsection{Code Obfuscation}
Code obfuscation refers to the process of leveraging transformations to make functionally equivalent programs that are difficult to understand, aiming to protect the intellectual property of developed software or hide malicious behaviors. Such transformations include encoding data, opaque predicates~\cite{xu2018manufacturing}, flattening control flow~\cite{laszlo2009obfuscating}, etc. For example, \cite{liu2021software} proposed a Mixed Boolean-Arithmetic (MBA) expression to mix the bitwise operations (e.g., AND, OR, and NOT) and arithmetic operations (e.g., ADD and IMUL), thereby creating more difficulties for attackers to analyze programs. In modern software development, obfuscation has been used to protect critical code parts against reverse engineering \cite{eidreverse}. However, the significant progress of LLMs challenges existing code obfuscation-based protection approaches, questioning whether existing obfuscation can still be effective against LLMs-based de-obfuscation and preserve sensitive information in developed software. Hence, it becomes urgent to evaluate the code analysis results of LLMs when code is obfuscated. In this work, we leverage an open-source JavaScript obfuscation tool~\cite{jsobtool} as well as a state-of-the-art obfuscator~\cite{romano2022wobfuscator} to generate obfuscated code and investigate whether LLMs are able to understand their functionality. As presented in Listing~\ref{noobf} and Listing~\ref{withobf}, a simple "Hello World" function written in JavaScript can be changed into an unintelligible one.
\begin{lstlisting}[language=java, caption= No obfuscation., breaklines=true, numbers=left, label=noobf, basicstyle=\tiny]
function hi() {
  console.log("Hello World!");
}
hi();
\end{lstlisting}
\begin{lstlisting}[language=java, caption= After obfuscation by \cite{jsobtool}.,breaklines=true, numbers=left, label=withobf, basicstyle=\tiny]
function _0x1ec3(){var _0x3ed452=['259790KgLPlj','297688NTFutg','35ACWDkX','145716kEyGyf','18SFCPKB','1701952aKOEga','192jjwxUU','5LPjNwr','142417rtWDUq','Hello\x20World!','121610lhBPGW','2032200UghFpX','5nCOmEq','log'];_0x1ec3=function(){return _0x3ed452;};return _0x1ec3();}(function(_0x22b342,_0x360ffb){var _0x5047be=_0xfb3c,_0x4c7c5c=_0x22b342();while(!![]){try{var _0x40c3be=parseInt(_0x5047be(0x90))/0x1*(-parseInt(_0x5047be(0x8e))/0x2)+-parseInt(_0x5047be(0x95))/0x3+parseInt(_0x5047be(0x97))/0x4*(parseInt(_0x5047be(0x99))/0x5)+parseInt(_0x5047be(0x8f))/0x6+parseInt(_0x5047be(0x94))/0x7*(parseInt(_0x5047be(0x93))/0x8)+parseInt(_0x5047be(0x96))/0x9*(-parseInt(_0x5047be(0x92))/0xa)+parseInt(_0x5047be(0x9a))/0xb*(-parseInt(_0x5047be(0x98))/0xc);if(_0x40c3be===_0x360ffb)break;else _0x4c7c5c['push'](_0x4c7c5c['shift']());}catch(_0x33f4b4){_0x4c7c5c['push'](_0x4c7c5c['shift']());}}}(_0x1ec3,0x52a68));function _0xfb3c(_0x257a0b,_0x17c420){var _0x1ec321=_0x1ec3();return _0xfb3c=function(_0xfb3ca7,_0x44b6b2){_0xfb3ca7=_0xfb3ca7-0x8d;var _0x34ca8b=_0x1ec321[_0xfb3ca7];return _0x34ca8b;},_0xfb3c(_0x257a0b,_0x17c420);}function hi(){var _0x2da467=_0xfb3c;console[_0x2da467(0x91)](_0x2da467(0x8d));}hi();
\end{lstlisting}
\section{Experiment Settings}\label{SecEvalSet}
We first conduct a systematic analysis of the ability of LLMs to perform code analysis-related tasks. In this section, we introduce the models and datasets used in our experiments.

\subsection{LLM Selection}
In our analysis, we select five representative popular LLM models to deploy:
\begin{itemize}
    \item GPT-3.5-turbo: GPT-3.5 is a set of LLMs offered by OpenAI, and it is the default model that is used for the popular LLM web tool ChatGPT~\cite{chatgpt}.
    \item GPT-4: GPT-4 is improved based on GPT-3.5 and it has been reported to be able to handle different types of tasks~\cite{bubeck2023sparks}. It is also one of the most advanced general-purpose LLMs currently.
    \item LLaMA-2-13B~\cite{touvron2023llama}: LLaMA is a set of foundation LLMs offered by Meta, with model parameter sizes ranging from 7B to 70B. LLaMA-2-13B contains 13B parameters and is reported to output a lot of larger models. We select LLaMA-2-13B as a representative medium-size LLM.
    \item Code-LLaMA-2-13B-Instruct~\cite{roziere2023code}: Code LLaMA is a family of LLMs provided by Meta, using the previously mentioned LLaMA family as foundation models. Code-LLaMAs are fine-tuned on code data, and it has been reported to outperform other public models targeting code-related tasks.
    \item StarChat-Beta~\cite{Tunstall2023starchat-alpha}: StarChat-Beta is an open-source model trained for assisting coding tasks. It has 16B parameters and is capable of handling multiple programming languages.
\end{itemize}
We select these models for the reason that they are widely used and are the current state-of-the-art publicly-available LLMs with the capability to perform code tasks.

\subsection{Prompt Construction}
We interact with the selected LLMs in different steps in our experiments, including instructing the LLMs to analyze code and other measurement process (discussed later in this section). All the prompts constructed in this paper either involve simple instructions (``\textit{Analyze the code and tell me what it does.}'', etc.) or adhere to empirically proven structures, such as assigning roles~\cite{jessen2023chit,yuan2023evaluating}. Specific prompts we utilize are provided in Section~\ref{SecEvalSet} and Section~\ref{SecCaseStudy}.

\subsection{Non-Obfuscated Code Dataset}
In this study, we aim to systematically evaluate the ability of LLMs to analyze, comprehend, and summarize code. We first construct a non-obfuscated source code dataset. Three languages are selected in this phase of study: JavaScript, Python, and C. We select JavaScript and Python since they are ranked the top $2$ most used languages on Github~\cite{githubtop} and we use them as the representatives of scripting languages. We select C as it is also ranked high (\#9) and it can be the representative of lower-level programming languages. All three languages we select are widely deployed over the Internet and in various computing systems. 

For JavaScript, we employ the combination of:
\begin{itemize}
    \item The Octane 2.0 benchmark~\cite{octane}, which is a JavaScript benchmark to test JavaScript performance. It contains benchmarks to test typical functionalities of JavaScript users.
    \item A list of practical JavaScript applications~\cite{jsapps}, including password generator, etc.
\end{itemize}

For Python, we use the Python branch of the CodeSearchNet dataset~\cite{husain2019codesearchnet} provided by Google. It contains samples of Python projects crawled from the Internet.

For C, we utilize the combination of:
\begin{itemize}
    \item A list of classic performance benchmarks, including CoreMark~\cite{coremark}, Dhrystone~\cite{weicker1984dhrystone}, Hint Benchmark~\cite{gustafson1995hint}, Linpack~\cite{dongarra2003linpack}, NBench~\cite{nbench}, Stream Benchmark~\cite{mccalpin1995memory}, TripForce~\cite{tripforce} and Whetstone~\cite{curnow1976synthetic}.
    \item A subset of the POJ-104 dataset~\cite{mou2016convolutional,poj104}, which consists of C code samples to solve $104$ different programming problems in an OJ system. The POJ-104 dataset provides multiple different solutions for each programming problem. In this study, for each programming problem, we randomly select one file from the POJ-104 dataset to form the dataset used in this work.
\end{itemize}
 
For each code file in our dataset, we develop scripts to automatically remove comments to eliminate their impacts on analysis results. Our goal is to let LLMs focus on code, without providing unnecessary natural language hints in code comments.

\begin{figure*}[htbp!]
    \centering
    \begin{subfigure}[t]{0.24\linewidth}
         \centering
         \includegraphics[width=\linewidth]{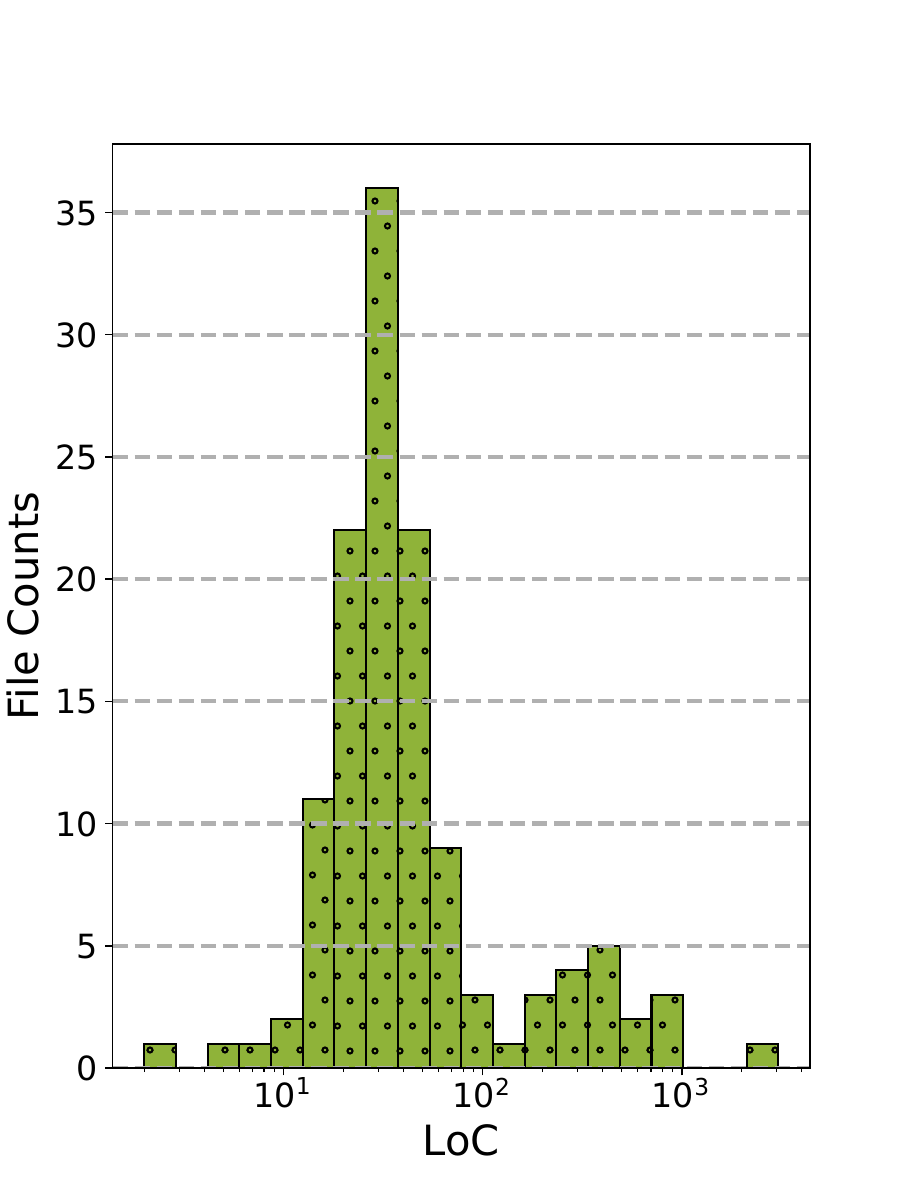}
         \caption{C.}
    \end{subfigure}
    \begin{subfigure}[t]{0.24\linewidth}
         \centering
         \includegraphics[width=\linewidth]{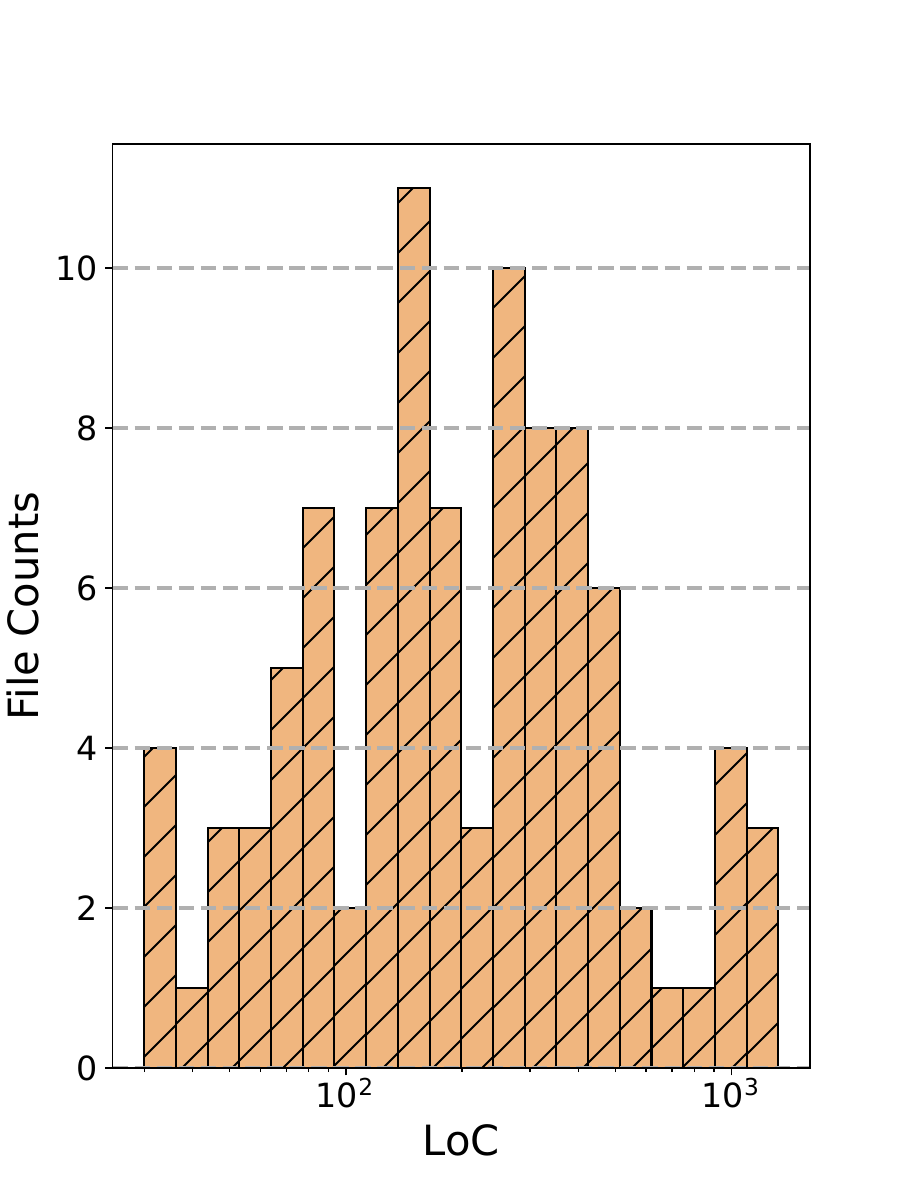}
         \caption{Python.}
    \end{subfigure}
    \begin{subfigure}[t]{0.24\linewidth}
         \centering
         \includegraphics[width=\linewidth]{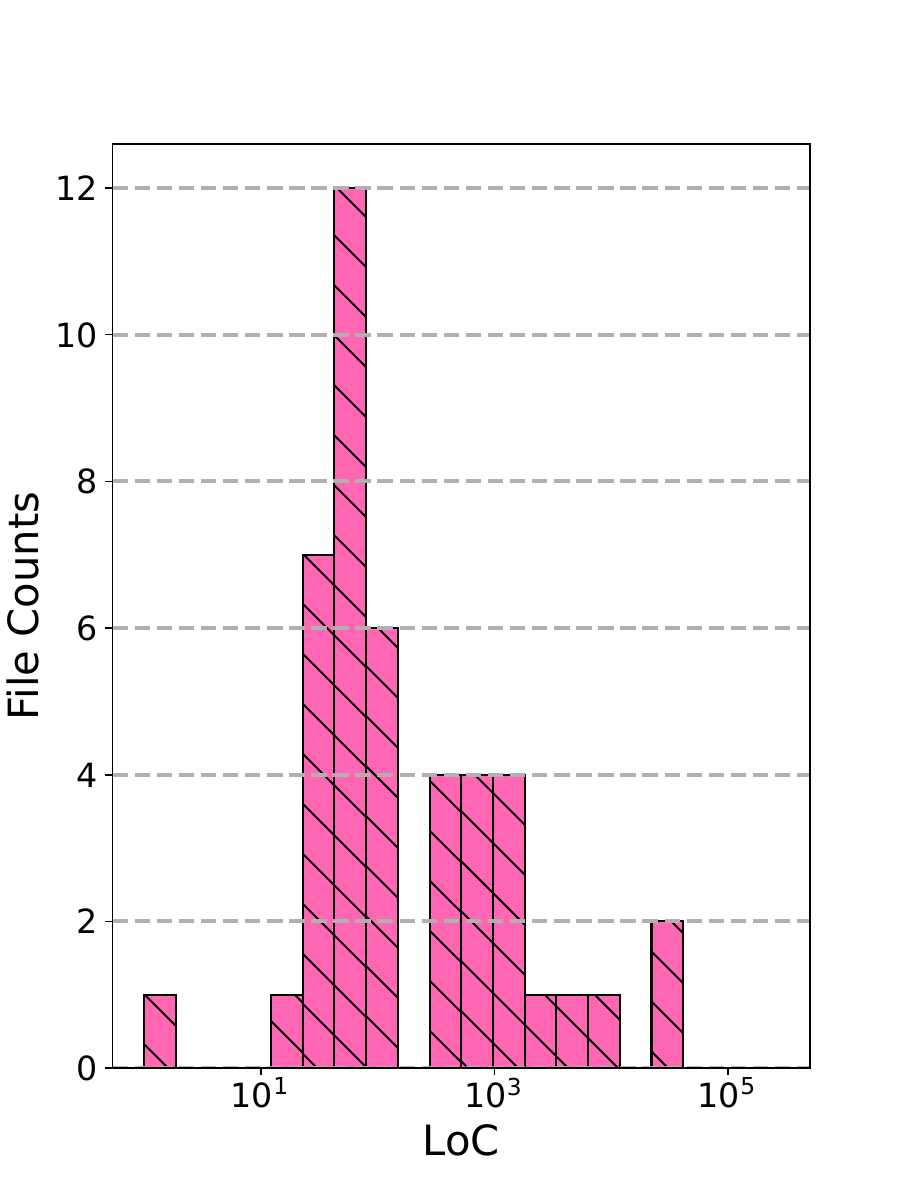}
         \caption{JavaScript.}
    \end{subfigure}
    \begin{subfigure}[t]{0.24\linewidth}
         \centering
         \includegraphics[width=\linewidth]{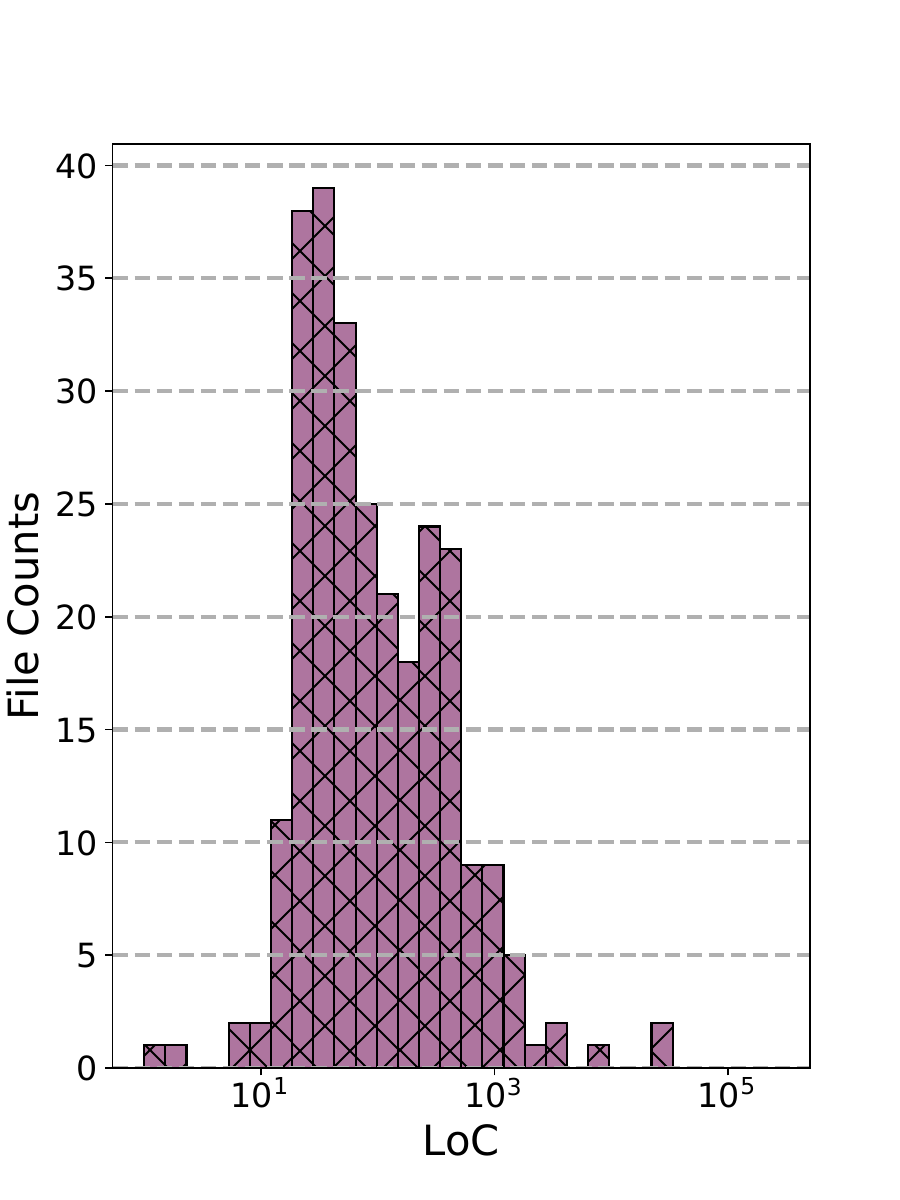}
         \caption{Overall.}
    \end{subfigure}
    \caption{Histograms regarding LoC statistics of our non-obfuscated source code dataset.}
    \label{FigHist}
\end{figure*}

The histograms regarding the line of code (LoC) distributions of processed files are shown in Figure~\ref{FigHist}. We can see that our dataset includes code samples spanning from very short (only a few lines) to very large-scale (over $10$k lines). Besides, the code samples in our dataset are from different sources covering different use cases. We believe the coverage of this dataset is sufficient, since we have chosen the most popular programming languages and selected a diverse set of code samples of different scales from different application scenarios.

\subsection{Obfuscated Code Dataset}
We choose to perform obfuscation on the JavaScript branch of our non-obfuscated code dataset. This choice was driven by the prevalence of code obfuscation practices in the JavaScript language, since JavaScript code is usually visible to web users, making additional obfuscation protection necessary. Besides, malicious JavaScript developers also apply obfuscation techniques to their code to hide the actual intent of their scripts. We use: ($1$) an open-source tool called JavaScript Obfuscator~\cite{jsobtool} to generate the obfuscation version of our JavaScript code; ($2$) Wobfuscator~\cite{romano2022wobfuscator}, a state-of-the-art obfuscator that transforms part of the code to WebAssembly~\cite{haas2017bringing}. The tested obfuscation methods are listed below:
\begin{enumerate}
    \item Default obfuscation (DE), which replaces identifier names with meaningless randomly generated strings, simplifies source code to reduce readability, placing strings in separate arrays, etc.~\cite{jsobtool}
    \item Dead code injection (DCI), which inserts random unrelated code blocks to the source code~\cite{christodorescu2003static} in addition to the default scheme.
    \item Control flow flattening (CFF), which transforms the structure of a program and hides control flow information~\cite{laszlo2009obfuscating} in addition to the default scheme.
    \item Split string (SS), which splits long strings into shorter chunks in addition to the default scheme to alleviate information leakage from embedded texts~\cite{xu2012power}. 
    \item Wobfuscator (WSM)~\cite{romano2022wobfuscator}, which performs cross-language obfuscation to the provided code.
\end{enumerate}
Our chosen obfuscation methods cover classic code obfuscation techniques (the first $4$) and a more recently developed obfuscation tool (Wobfuscator). By testing the performance of LLMs on these obfuscated code samples, we will be able to understand how these obfuscation techniques impact the ability of LLMs to understand code.

Besides obfuscating our previously acquired source code, we also combine existing obfuscated code from online resources. We integrate winner code samples from the International Obfuscated C Code Contest (IOCCC)~\cite{ioccc}, which is a contest that challenges participants to write the most obscure and confusing C code. Instead of asking LLMs to explain the code, we consider adding an extra challenge to generate de-obfuscated code and see if the generated code can be compiled and run. Experiments on this part of our obfuscated code dataset evaluate the performance of LLMs when facing more flexible and non-standard obfuscation techniques.

\begin{figure}[htbp]
    \centering
    \includegraphics[width=.95\linewidth]{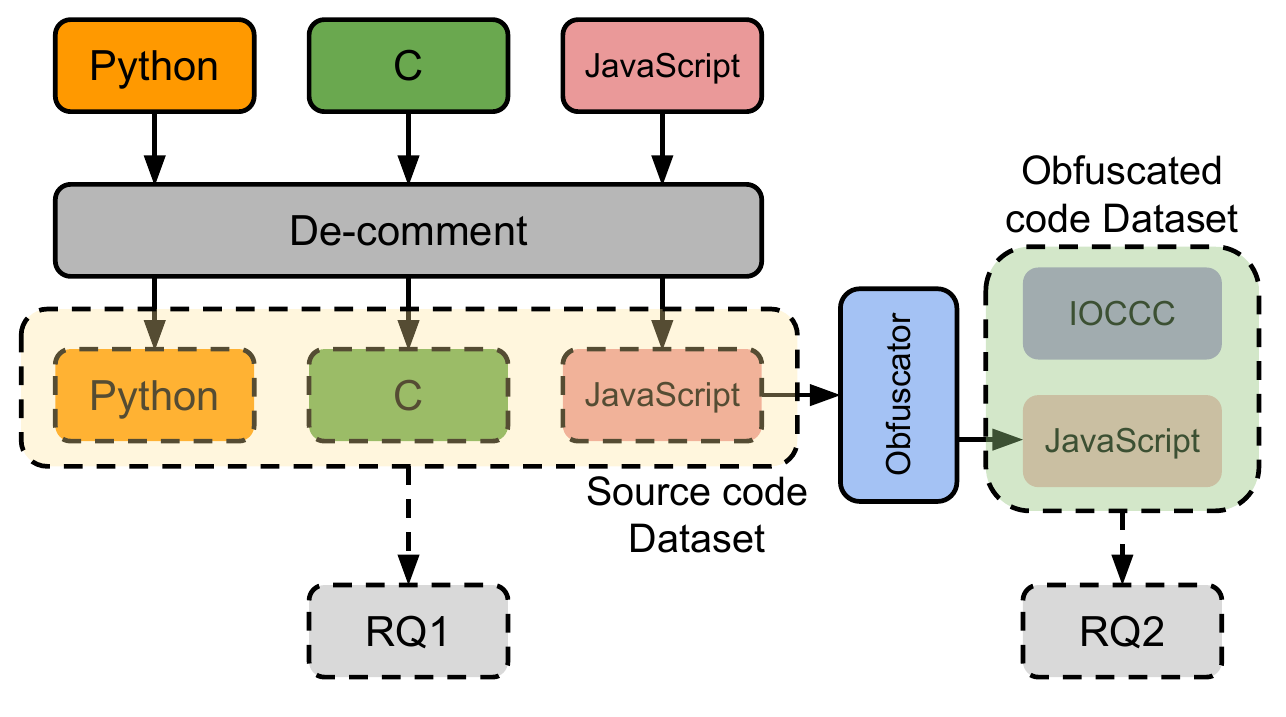}
    \caption{Diagram of our experiment pipeline.}
    \label{FigAnalysisDiagram}
\end{figure}

\subsection{Measurement Method}
After collecting the response of code analysis from our target LLMs, we start a manual validation process to check the correctness of the analysis results. Since all the source code files we employ only come with a maximum of a few tens of lines of concise comments and the description labels attached (if there are any) are also succinct, we can not rely on directly applying certain quantitative metrics (e.g., n-gram~\cite{brown1992class}) to determine the correctness of generated analysis results, which necessitates a manual validation process.


\textbf{Ground Truth.} We choose to first manually examine the outcomes of GPT-4~\cite{sanderson2023gpt}, since it has better natural language generation and deduction capability. During this manual validation process, four graduate/PhD-level students majoring in computer science or electrical engineering, each with over 5 years of programming experience, read the code and assess GPT-4's generated explanation for accuracy and potential errors. The generated explanation for each code file is labeled `correct' if the functionality of the code and the description match. Cross-checking is conducted among the students to minimize biases. After this step, we consider those descriptions marked `correct' as the ground truth and use those descriptions for further comparison among different LLMs.

\textbf{Comparison Metrics.} When evaluating the generated explanation (either non-obfuscated or obfuscated), we utilize the following metrics: 
\begin{enumerate}
    \item Cosine similarity (ranging from $0$-$1$), since it is widely used in natural language processing-related tasks and can serve as a coarse grain metric in this study; 
    \item A Bert-based semantic similarity score~\cite{semantictextsimilarity} (ranging from $0$ to $5$) that is more advanced in comparing the similarity of natural language inputs;
    \item ChatGPT-based evaluation~\cite{zheng2023judging,chiang2023can,yuan2023evaluating,bubeck2023sparks} (\textbf{True} or \textbf{False}). In our evaluation, GPT-4 model will receive the following instructions first:
    \begin{description}
        \item[Instruction:]``\textit{You are given two descriptions of two code snippets: Description 1 and Description 2. Corresponding code snippets are not available. From the given text, do you think the two descriptions correspond to two code snippets with roughly similar functionalities? Output should be "Yes" if similar, or "No" otherwise, followed by a brief justification of how this is determined.}''
    \end{description}
    We then use the generated output to examine the correctness of each model.
\end{enumerate}

\section{Results}\label{SecResult}
In this part, we present our results on our non-obfuscated code dataset and obfuscated code dataset, mainly to answer the two research questions raised in Section \ref{SecIntro}:
\begin{itemize}
    \item \textbf{RQ1}: Do LLMs understand source code? (Section~\ref{SecResultRQ1})
    \item \textbf{RQ2}: Can LLMs comprehend obfuscated code? (Section~\ref{SecResultRQ2})
\end{itemize}

LLMs are prompted 
\begin{description}
\item[] ``\textit{Analyze the code and tell me what it does.}''
\end{description}
to perform the analysis tasks. We also present findings in our experiments.

\subsection{Results on Non-Obfuscated Code Dataset}\label{SecResultRQ1}
\begin{figure}[htbp]
    \centering
    \includegraphics[width=\linewidth]{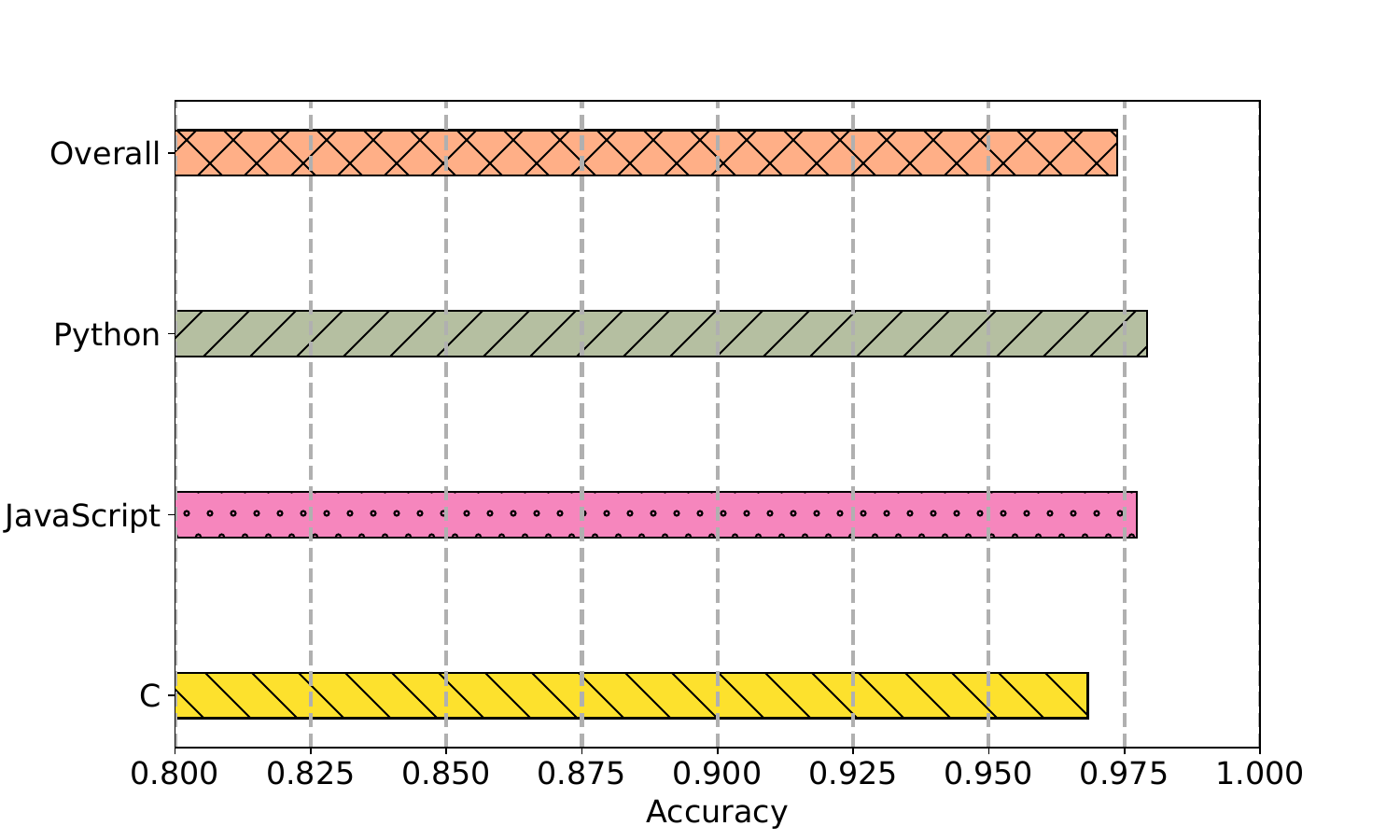}
    \caption{GPT-4 accuracy.}
    \label{FigGroundTruthAccu}
\end{figure}

\textbf{GPT-4 Results.} The manually verified accuracy results of GPT-4 are shown in Figure~\ref{FigGroundTruthAccu}. We can see from Figure~\ref{FigGroundTruthAccu} that the accuracy performance of GPT-4 on the three languages (C, JavaScript, and Python) is high. For all of the three languages, over $95\%$ of the analysis generated by GPT-4 aligns with the actual content of the code samples. The overall accuracy rate is $97.4\%$, indicating that GPT-4 can serve as a powerful code analysis tool.
After meticulously analyzing the generated outputs of GPT-4, we also have the following findings.

\begin{mdframed}[backgroundcolor=gray!20]
\noindent
    \textbf{Finding 1}: GPT-4 is able to recognize code snippets from popular open-source software repositories. 
\end{mdframed}

It has been reported that LLMs can leak memorized information acquired from training during conversations~\cite{carlini2021extracting}. In our experiments, we observe this phenomenon multiple times. For example, our dataset contains source code files from Heron project~\cite{heron}, a real-time analytics system developed by Twitter. Even though the source code has been de-commented and does not contain the keyword ``Twitter'', GPT-4 still successfully connects the source code with Twitter by stating: \textit{``The code is written in Python and is part of the Heron project, a real-time stream processing framework developed by Twitter.''} at the beginning of its response. In another example, when we feed GPT-4 with a local JQuery copy from the Octane benchmark set, GPT-4 also successfully recognizes that the provided code is from the JQuery library:\textit{``The provided code is a part of the jQuery library, ...''} at the beginning of its response. The same phenomenon is also observed in the responses from other models, e.g. GPT-3.5.

This indicates that the training data of these models contain code samples from popular open-source repositories, and LLMs are able to connect the provided code snippets with their memorized information. This might be helpful when analyzing code, since being able to connect source code with existing software implementation can accelerate the process of understanding the intent of target code snippets. Please note that this finding does not suggest that we are conducting our experiments in a ``test on training set'' manner. In our experiments, we focus on examining the generated detailed explanation of each function/code sample, which is more likely deducted from the source code and unlikely to be contained in the training set. Our experiments on newer repositories in Section~\ref{SecCaseStudy} also indicate that LLMs do not need to use memorized information to assist code analysis.

\begin{mdframed}[backgroundcolor=gray!20]
\noindent
\textbf{Finding 2}: GPT-4 occasionally makes wrong associations.
\end{mdframed}

In our experiments, though GPT-4 has a high analysis success rate, there are still a few code samples that GPT-4 cannot correctly analyze. When using GPT-4 for analyzing a Python file related to stock market utilities, GPT-4 incorrectly concludes that the code uses the \texttt{pandas} package, even though it is not imported in the code sample. We believe this is caused by \texttt{matplotlib} and \texttt{numpy} packages being imported. These packages usually appear together in data analytics scripts that are likely part of the training set. GPT-4 possibly makes an incorrect association based on its memorized contents.

\begin{mdframed}[backgroundcolor=gray!20]
\noindent
\textbf{Finding 3}: GPT utilizes information provided in identifier names to assist code analysis.
\end{mdframed}

It is intriguing to find that as a language model, GPT-4 utilizes natural language hints embedded in identifier names to assist code analysis like humans. For example, during the analysis for the CoreMark benchmark, GPT-4 recognizes identifiers named ``\texttt{ee\_u8}'', ``\texttt{ee\_u16}'', etc. are custom data types and concludes that the underlying mapping may vary across different embedded systems. When analyzing the JavaScript Octane benchmark, GPT-4 successfully recognizes that the code samples are used for benchmarking/testing simply from the appearance of a variable name ``\texttt{Benchmark}''. This suggests that the LLMs will be able to generate higher-quality analysis and provide more information on clearly written code.

\begin{figure}[ht!]
    \centering
    \begin{subfigure}[t]{\linewidth}
         \centering
         \includegraphics[width=.9\linewidth]{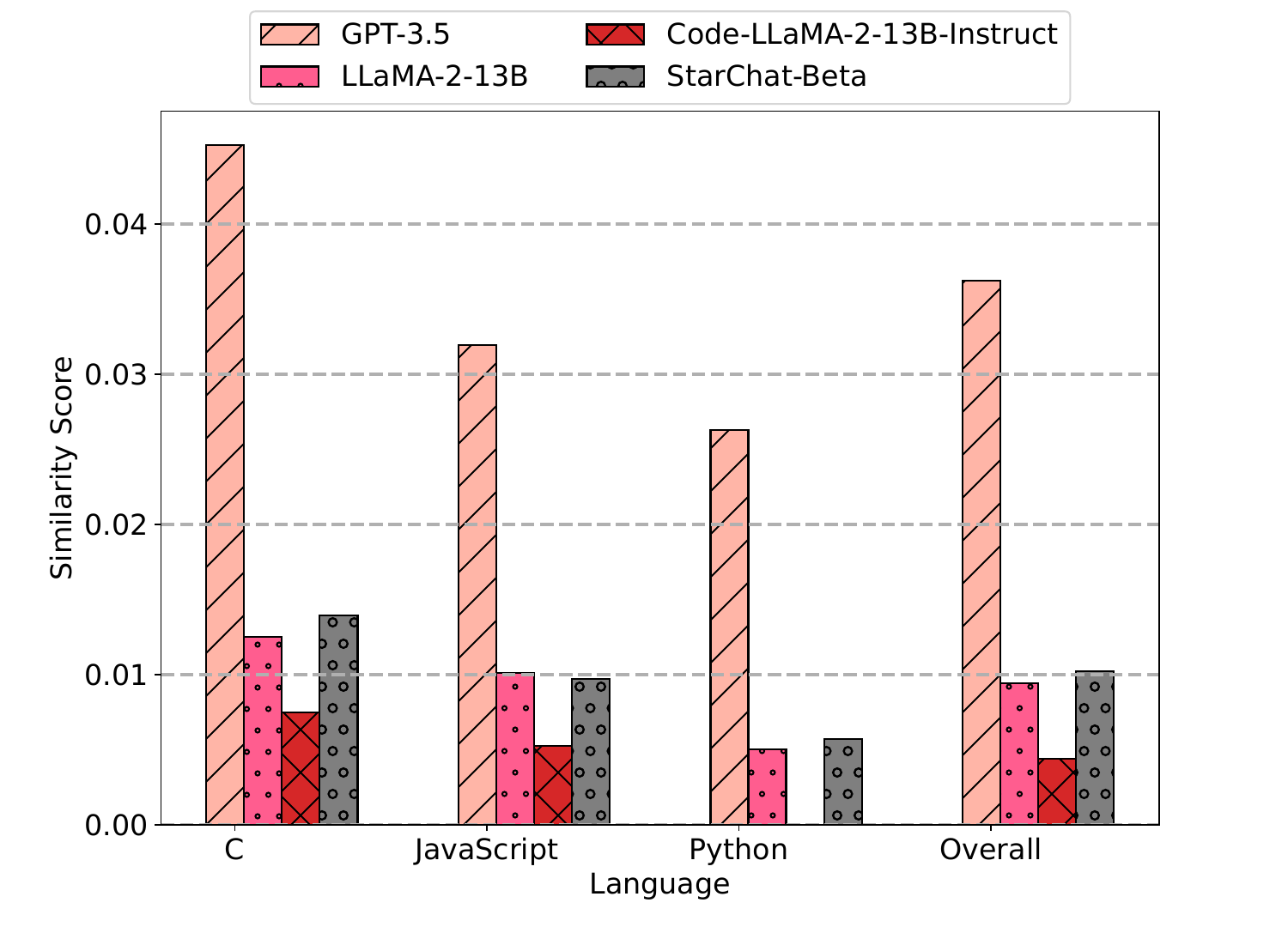}
         \caption{Cosine similarity score.}
    \end{subfigure}
    \begin{subfigure}[t]{\linewidth}
         \centering
         \includegraphics[width=.9\linewidth]{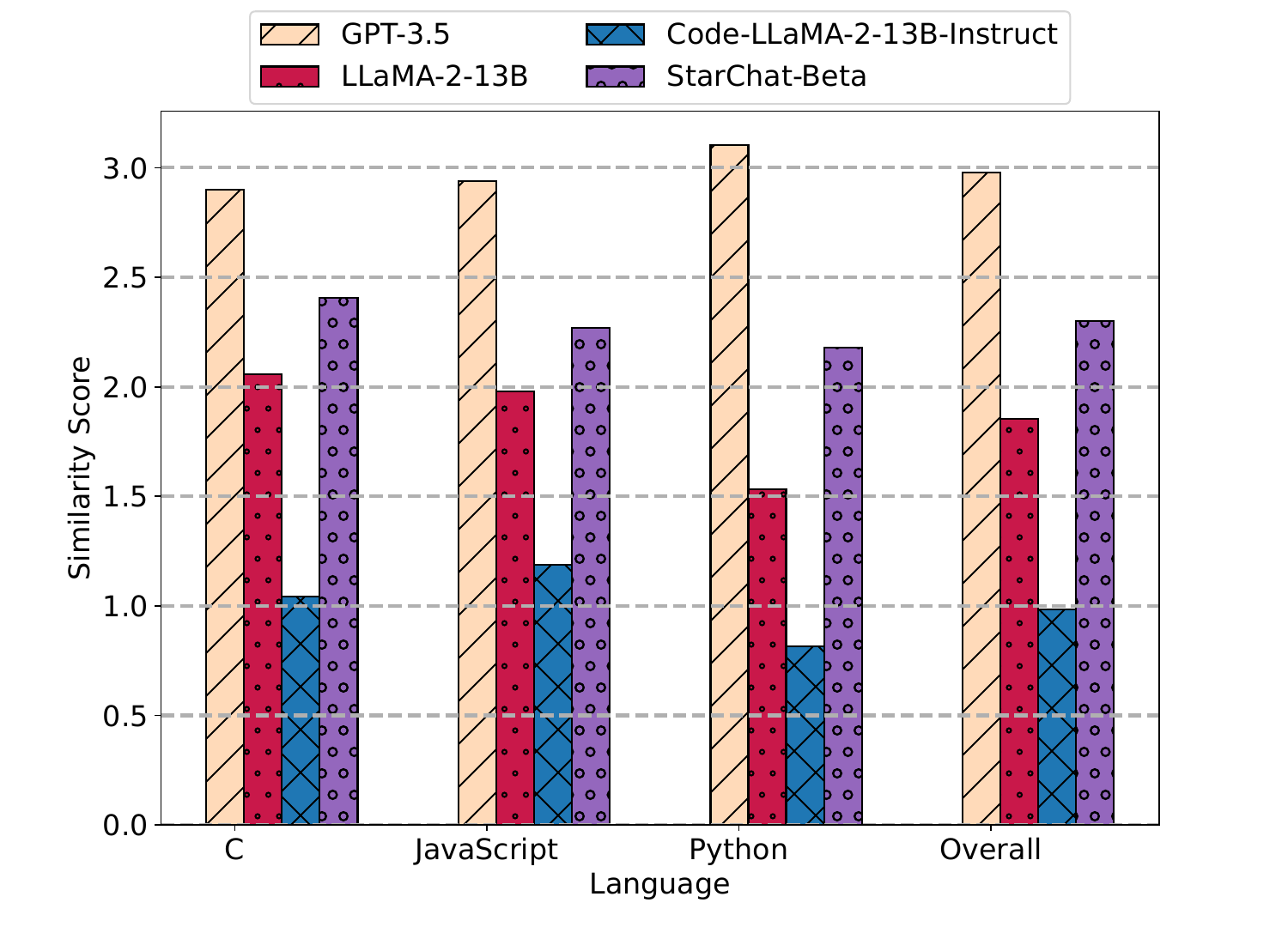}
         \caption{Bert-based semantic similarity scores, with the evaluator trained on web data.}
    \end{subfigure}
    \begin{subfigure}[t]{\linewidth}
         \centering
         \includegraphics[width=.9\linewidth]{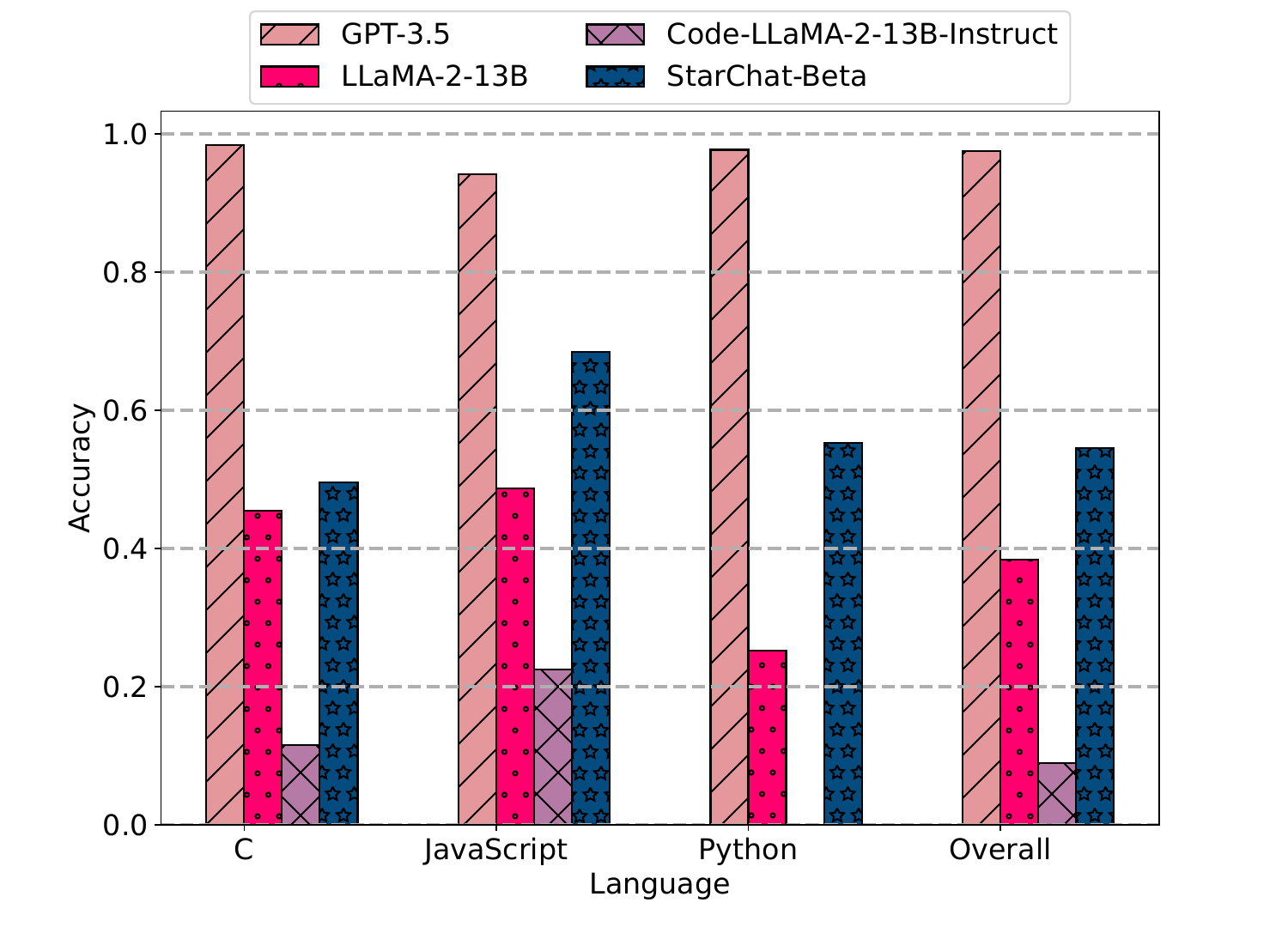}
         \caption{ChatGPT measured accuracy results.}
    \end{subfigure}
    \caption{Similarity score/accuracy of analysis results generated by different models. For LLaMA series, results are obtained after we manually extract the natural language contents from the generated outputs.}
    \label{FigCodeSim}
\end{figure}

\textbf{Analysis Accuracy of Other Selected Models.} In Figure~\ref{FigCodeSim}, we present the cosine similarity score [Figure~\ref{FigCodeSim}~(a)], Bert-based semantic similarity [Figure~\ref{FigCodeSim}~(b)] and accuracy of code explanation measured by GPT-4 [Figure~\ref{FigCodeSim}~(c)]. We can see that GPT-3.5 achieves similar performance as GPT-4, indicating that both models can achieve high performance in code analysis tasks. However, for the LLaMA-series models and StarChat-Beta, the accuracy performance is significantly lower. More findings obtained by manually examining the generated results are provided below.

\begin{mdframed}[backgroundcolor=gray!20]
\noindent
\textbf{Finding 4}: Smaller models in our experiments (LLaMA-2-13B, Code-LLaMA-2-13B-Instruct, and StarChat-Beta) are unable to generate consistent paragraphs of code analysis results, unlike GPT-3.5 and GPT-4.
\end{mdframed}

While performing the designated code analysis tasks, instead of generating consistent paragraphs, we observe that LLaMA-2-13B tries to repeat questions and generates hints to itself. Rather than generating paragraphs to answer the question directly, all three models tend to generate multiple short and inconsistent statements. In many cases, LLaMA-2-13B rephrases the input question to a semantically different one (e.g. it rephrases \textit{``Analyze the following piece of code ''}  to \textit{``Please tell me what this code does and what are the vulnerabilities in this code.''}). When handling long code samples, StarChat-Beta, Code-LLaMA-2-13B-Instruct and LLaMA-2-13B tend to simply return part of the input code or return the re-written input code, without trying to provide explanations like GPT-3.5 and GPT-4. These phenomena indicate that these models do not have the ability to process code analysis-related tasks. Results shown in Figure~\ref{FigCodeSim} are similarity results after we manually extract code contents from the generated outputs. Even after we remove the code contents, the performance of these models is still significantly lower than GPT-3.5.


\begin{mdframed}[backgroundcolor=blue!20]
\noindent
    \textbf{Answer to RQ1}:  For non-obfuscated code, larger models like GPT-3.5 or GPT-4 have a high probability of generating correct and detailed explanations for input code snippets, while the smaller models, even fine-tuned on code data, fail to generate correct outputs.
\end{mdframed}

\subsection{Results on Obfuscated Code Dataset}\label{SecResultRQ2}

\begin{figure}[htbp!]
    \centering
    \begin{subfigure}[t]{\linewidth}
         \centering
         \includegraphics[width=\linewidth]{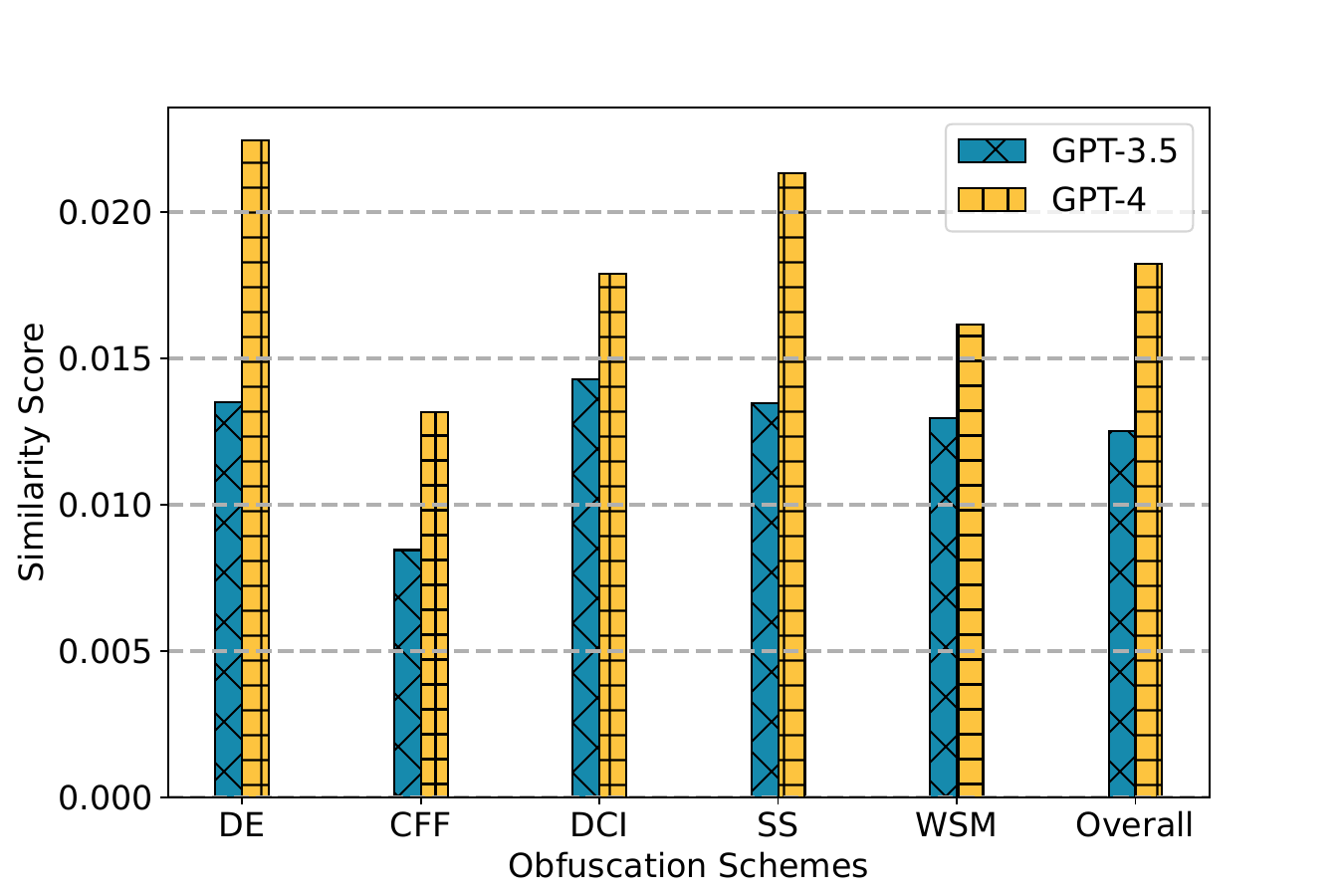}
         \caption{Cosine similarity score.}
    \end{subfigure}
    \begin{subfigure}[t]{\linewidth}
         \centering
         \includegraphics[width=\linewidth]{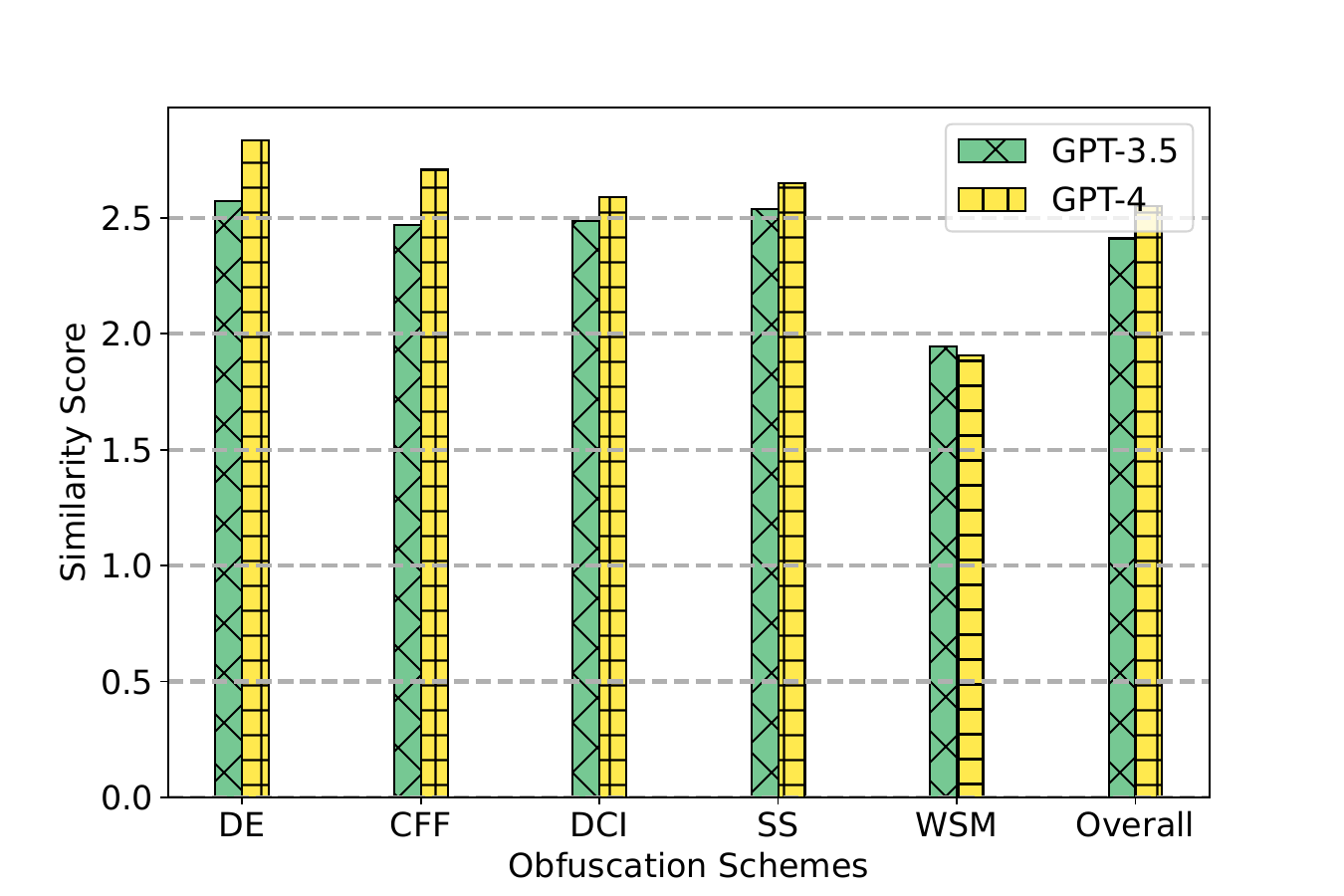}
         \caption{Bert-based semantic similarity scores, with the evaluator trained on web data.}
    \end{subfigure}
    \begin{subfigure}[t]{\linewidth}
         \centering
         \includegraphics[width=\linewidth]{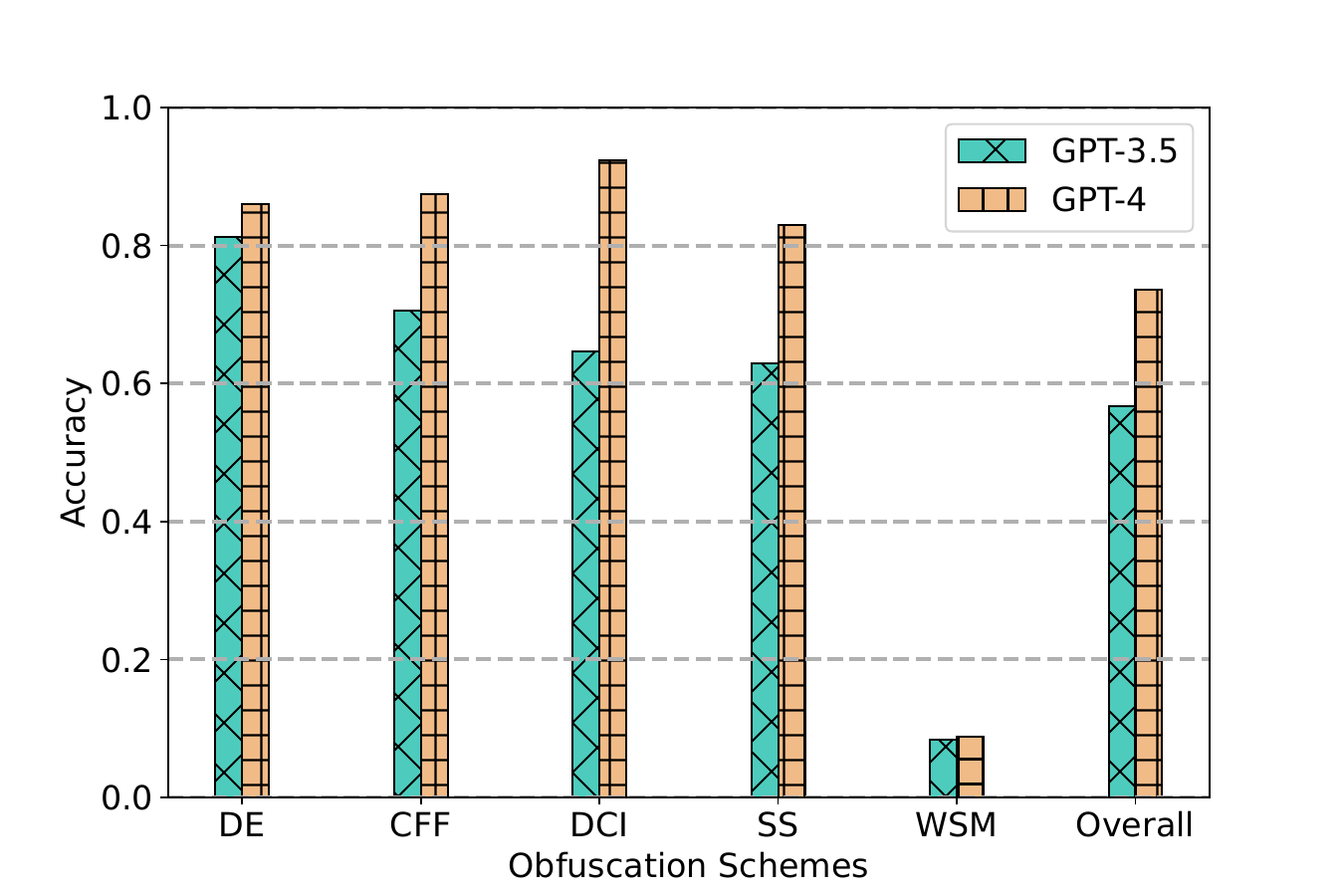}
         \caption{ChatGPT measured accuracy results.}
    \end{subfigure}
    \caption{Similarity score/accuracy of obfuscated code analysis results generated by different models.}
    \label{FigJSObfSim}
\end{figure}

\textbf{Obfuscated Code Analysis Capability Evaluation.}
We first focus on analyzing the obfuscated JavaScript code samples. We use the same prompt to instruct LLMs to analyze code and generate explanations as in the previous part. Although during the obfuscation process, function blocks are rewritten in different ways, the functionalities will not change and we observe that the outputs of LLMs are of very similar formats. Therefore, we will keep using the same set of metrics to evaluate the effectiveness of code explanation. Results regarding similarity scores and accuracy of obfuscated code analysis are shown in Figure~\ref{FigJSObfSim}.


From Figure~\ref{FigJSObfSim}~(c), we can see that GPT-4 still shows exceptional analysis accuracy, reaching $87\%$ accuracy. GPT-3.5, on the other hand, is impacted by obfuscation techniques, especially when more advanced techniques are applied (control flow flattening, dead code injection, and split string). As for similarity score metrics shown in Figure~\ref{FigJSObfSim}~(a) and (b), we can see that both models suffer performance degradation, yet the similarity scores of GPT-4 are constantly higher than GPT-3.5. These results indicate that when facing obfuscated code analysis, GPT-4 is a better choice. There are other findings as presented below:

\begin{mdframed}[backgroundcolor=gray!20]
\noindent
    \textbf{Finding 1}: LLaMA-2-13B, Code-LLaMA-2-13B-Instruct and StarChat-Beta are unable to generate meaningful explanations once obfuscation techniques are applied.
\end{mdframed}

In our experiments, we find that all three models are unable to generate meaningful results. In most cases, they simply return part of the obfuscated code that is sent. Therefore, we do not include results from LLaMA-2-13B and Code-LLaMA-2-13B-Instruct in Figure~\ref{FigJSObfSim}. These results again indicate that these relatively small models do not possess the ability to perform code-analysis tasks properly.

\begin{mdframed}[backgroundcolor=gray!20]
\noindent
    \textbf{Finding 2}: Basic obfuscation techniques (such as DE in our paper) only slightly influence the ability of GPT models to perform code analysis.
\end{mdframed}

In our experiments, we found that basic obfuscation techniques like replacing identifier names do not significantly impair the ability of GPT models to produce analysis results. Indeed, both models lose information contained in identifier names, but they are still able to extract sufficient information from the execution flow and remaining strings and provide relatively reliable explanations.

\begin{mdframed}[backgroundcolor=gray!20]
\noindent
    \textbf{Finding 3}: LLMs are not able to decipher obfuscated code generated by Wobfuscator.
\end{mdframed}

From Figure~\ref{FigJSObfSim}, we can see that the application of Wobfuscator significantly reduces the accuracy of generated code explanations. By analyzing the generated outputs, we observe that the insertion of WebAssembly code severely hinders the understanding of source code, especially as reflected by the Bert-based similarity score [Figure~\ref{FigJSObfSim}~(b)] and ChatGPT-based score [Figure~\ref{FigJSObfSim}~(c)]. The two GPT models do not decipher the inserted WebAssembly code and hence fail to properly understand the functionality of the provided code.

\begin{mdframed}[backgroundcolor=gray!20]
\noindent
    \textbf{Finding 4}: The ability of GPT models to decipher longer and more complicated obfuscated code is limited.
\end{mdframed}

We observe that GPT models perform worse when facing longer and more complicated code, which is expected. In our experiments, most code explanations that are classified as wrong are generated from longer code samples. Both GPT-3.5 and GPT-4 are able to maintain high accuracy when processing shorter and simpler website javascript applications. However, when facing larger code samples with more complicated functionalities, the generated responses are more error-prone and sentences like ``\textit{More information is needed}'' appear more frequently, especially for GPT-3.5.

\textbf{De-Obfuscated Code Generation.} Besides comparing code explanation results on the obfuscated JavaScript dataset, we also test the ability of LLMs to generate de-obfuscated code. Since it is a challenging task, we only perform this evaluation on the two most powerful models we have, i.e. GPT-3.5 and GPT-4. We select $100$ contest winner projects after the year 2011 competition of IOCCC~\cite{ioccc} and instruct LLMs to perform code de-obfuscation tasks. LLMs are prompted:
\begin{description}
    \item[] \textit{``You are an expert in code analysis. De-obfuscate the code and generate a readable new version.''}
\end{description}
and we directly feed the original code to LLMs. We select IOCCC because a more diverse and flexible set of obfuscation techniques are applied, compared to our automatically generated JavaScript dataset.

We evaluate the ability of LLM to generate de-obfuscated code from the following $3$ aspects:
\begin{enumerate}
    \item Whether code can be generated;
    \item (If generated) whether the generated code can pass compilation;
    \item (If compilable) whether the compiled code produces correct outputs.
\end{enumerate}
Corresponding statistical results are presented in Figure~\ref{FigDeobfSucc}. Our findings are presented below.

\begin{figure}[htbp]
    \centering
    \includegraphics[width=\linewidth]{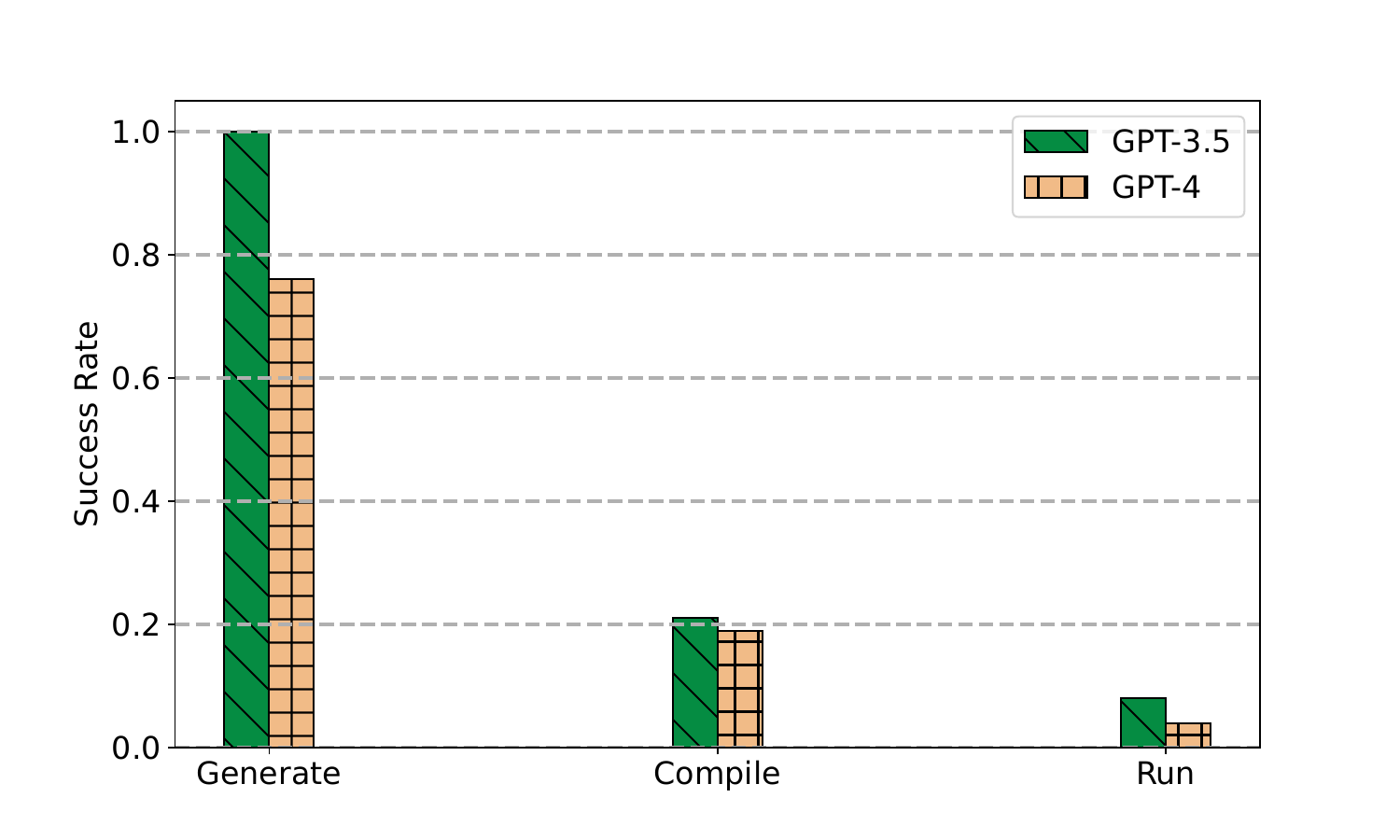}
    \caption{Success rate of different models regarding de-obfuscation tasks.}
    \label{FigDeobfSucc}
\end{figure}

\begin{mdframed}[backgroundcolor=gray!20]
\noindent
    \textbf{Finding 5}: Both GPT models in our evaluation fall short of generating compilable and runnable de-obfuscated code.
\end{mdframed}

As shown in Figure~\ref{FigDeobfSucc}, for GPT-3.5, although it successfully produces code outputs for all targets, only around $20\%$ of its output code samples are compilable, and only $8\%$ of generated code samples ($38\%$ of compiled code samples) can produce correct results. The performance of GPT-4 is worse than GPT-3.5. It is able to generate de-obfuscated code for only $76\%$ of experimented code samples. Only $19\%$ of all experimented code samples ($25\%$ of generated code) successfully pass the compilation step, and only $4\%$ of all experimented code samples ($21\%$ of compiled code) produce correct results. These statistics indicate that despite the strong ability to produce explanatory analysis results, their performance on de-obfuscated code generation is unsatisfying.

It is interesting to note that GPT-3.5 beats GPT-4 on these metrics in the code de-obfuscation task, despite the fact that it is a less advanced model. 

\begin{mdframed}[backgroundcolor=gray!20]
\noindent
    \textbf{Finding 6}: GPT-4 has a higher probability of associating given code samples to the IOCCC contest. However, details are incorrect.
\end{mdframed}

By examining responses generated by both models, we find that GPT-4 recognizes that $22$ of the provided code samples belong to the IOCCC contest, while GPT-3.5 is only able to identify $2$ of them. This indicates that despite IOCCC code samples being included in the training set of both models, GPT-4 is able to better associate the given input to its knowledge. However, when it tries to identify specific code samples, i.e., provide exact year and author names, the answers are wrong. For example, during our analysis, it identified a code sample that was awarded ``Best Handwriting in 2015'' as ``Best One-Liner in 2015''. We then asked \textit{``What is the Best One-Liner award receiver of IOCCC 2015?''}, and got a different but still incorrect answer that seems to be made up by GPT-4. This indicates that the given incorrect information is possibly not caused by misalignment in the training data.

\begin{mdframed}[backgroundcolor=gray!20]
\noindent
    \textbf{Finding 7}: GPT-4 has a higher probability of refusing to perform code-generation tasks.
\end{mdframed}

\begin{figure}[ht!]
    \centering
    \begin{subfigure}[t]{0.66\linewidth}
         \centering
         \includegraphics[width=\linewidth]{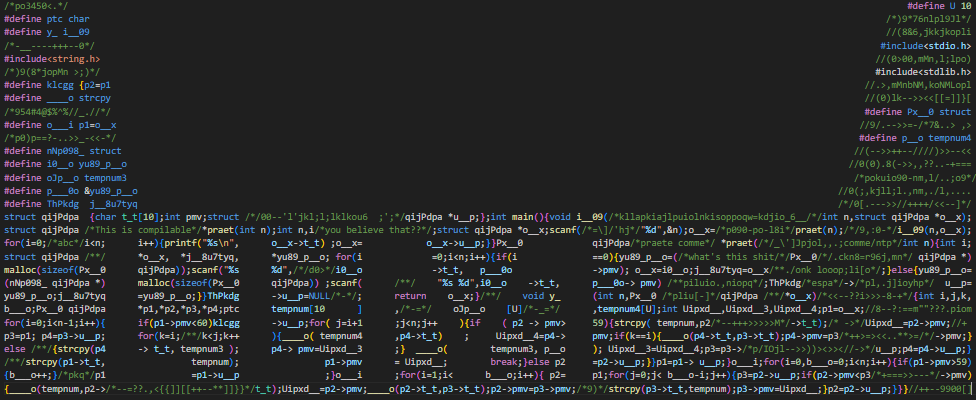}
         \caption{Reformatted code sample from POJ-104 dataset.}
    \end{subfigure}
    \begin{subfigure}[t]{0.34\linewidth}
         \centering
         \includegraphics[width=\linewidth]{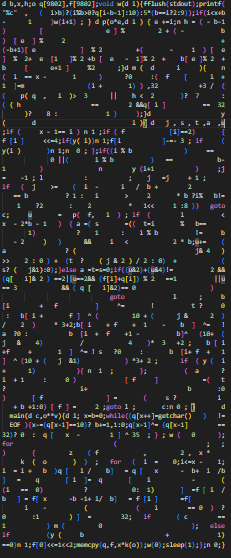}
         \caption{2011/blakely/blakely.c .}
    \end{subfigure}
    \begin{subfigure}[t]{0.355\linewidth}
         \centering
         \includegraphics[width=\linewidth]{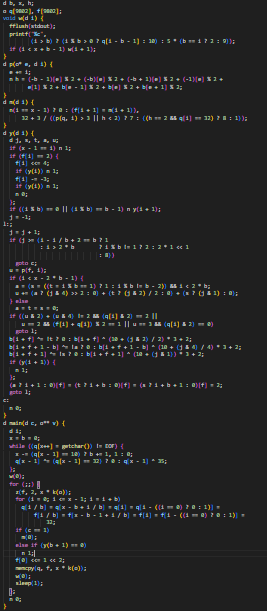}
         \caption{Reformatted 2011/blakely/blakely.c .}
    \end{subfigure}
    \caption{Reformatted code.}
    \label{FigReformat}
\end{figure}

As shown in Figure~\ref{FigDeobfSucc}, GPT-4's performance is significantly worse than GPT-3.5 in this task, especially at the code generation step. We observe that GPT-4 will admit that the code is obfuscated and requires lots of expert work to decipher, instead of generating code without additional remarks like GPT-3.5. By examining code samples that GPT-4 refuses to generate, we find that there are two factors that can potentially cause the failure of de-obfuscation:
\begin{enumerate}
    \item Code contains complicated logic.
    \item Code is formatted in a special style, unlike the traditional line-by-line organization.
\end{enumerate}
\begin{figure}[htbp!]
    \centering
    \begin{subfigure}[t]{0.215\linewidth}
         \centering
         \includegraphics[width=\linewidth]{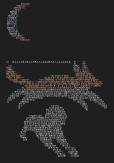}
         \caption{2013/cable2/prog.c .}
    \end{subfigure}
    \begin{subfigure}[t]{0.32\linewidth}
         \centering
         \includegraphics[width=\linewidth]{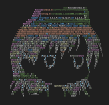}
         \caption{2013/misaka/misaka.c .}
    \end{subfigure}
    \begin{subfigure}[t]{0.35\linewidth}
         \centering
         \includegraphics[width=\linewidth]{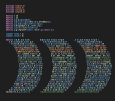}
         \caption{2019/giles/prog.c .}
    \end{subfigure}
    
    \caption{Samples of specially-formatted code.}
    \label{FigSpecialCode}
\end{figure}
Examples of specially formatted code snippets are shown in Figure~\ref{FigSpecialCode}. To rule out this factor, we select the \texttt{POJ-104/9/1618.c} sample code from the C branch of our non-obfuscated dataset and apply variable substitution and reformatting, resulting in a code snippet that is extremely hard to decipher with manual effort [Figure~\ref{FigReformat}~(a)]. We also select a file from the IOCCC dataset  (\texttt{blakely.c} from IOCCC 2011) and reformat the file to a normal format [Figure~\ref{FigReformat}~(b) and (c)].

In the first scenario, despite the obfuscation techniques applied, GPT-4 is still able to generate correct explanations as well as de-obfuscation results. In the second case, GPT-4 continues to refuse to generate de-obfuscated code. This leads to our next finding:

\begin{mdframed}[backgroundcolor=gray!20]
\noindent
    \textbf{Finding 8}: Text-level obfuscation does not influence the abilities of LLMs to perform de-obfuscation.
\end{mdframed}
This also aligns with our findings while conducting the analysis of the JavaScript code, indicating that employing complex logic is the only way to trick LLMs.

\begin{mdframed}[backgroundcolor=gray!20]
\noindent
    \textbf{Finding 9}: GPT-4 generates code with higher readability.
\end{mdframed}

We define readability as improved code formatting and more meaningful identifier names. Higher readability indicates superior code generation capability. Despite worse performance on code generation success rate, we find that code generated by GPT-4 has higher readability compared to code generated by GPT-3.5. GPT-4 is able to generate meaningful identifier names more often, while part of the GPT-3.5-generated code still seems obfuscated. This indicates that GPT-4 is still a better generative model if we take the quality of the generated code into consideration.

\begin{mdframed}[backgroundcolor=blue!20]
\noindent
    \textbf{Answer to RQ2}: Obfuscation techniques can impact the ability of LLMs to generate explanations. Smaller models in our experiments are unable to handle obfuscated code. GPT-3.5 and GPT-4 both drop in analysis accuracy, especially when facing Wobfuscator~\cite{romano2022wobfuscator}, though GPT-4 still has an acceptable and better accuracy performance on classic obfuscation methods. Without special optimization targeting de-obfuscated code generation, LLMs show a poor ability to generate functional de-obfuscated code.
\end{mdframed}

\section{Case Studies}\label{SecCaseStudy}
In this section, we conduct case studies and show how the capability of LLMs can be utilized for defensive static analysis. We first select two newly published Github repositories (one benign and one malicious) to test the performance of GPT-series models in malware analysis. We then select the Android \texttt{msg-stealer} virus and the WannaCry ransomware~\cite{national2017wannacry}  to further explore the performance of LLMs for analyzing decompiled and obfuscated code. Both viruses have been found in the real world. In both cases, code samples are directly obtained from decompilers. Decompiled and obfuscated code have a lot in common: both do not contain meaningful identifier names, and the control flow may not be straightforward. The complete responses of LLMs are contained in our online appendix.

\subsection{Github Repository Analysis}
In this case study, we select two repositories on Github: ($1$) \texttt{KratosKnife}~\cite{kratosknife}, which is a set of Python scripts of a botnet system; ($2$) \texttt{librarian}~\cite{librarian}, which is a Chrome extension for bookmark search. The reasons why we choose these two code repositories are: ($1$) With the comparison of malware and benign-ware, we can carefully observe the outputs and determine if any false alarms arise during the analysis process. ($2$) \texttt{librarian} is a new codebase (created 01/11/2024) that is guaranteed to be not included in GPT-4's training sets. Therefore, we can examine the ability of GPT-4 to analyze code without concerns for encountering any pre-learned patterns or memorization from GPT-4's training data. In the experiment, to generate explanations, we feed de-commented code files to GPT-4, provide file paths in the prompts, and instruct GPT-4 to analyze code. Responses of the previous file is passed as an assistant message, so that the whole analysis process consists of continued sessions. After this, we feed the generated results one at a time to GPT-4 and ask ``\textit{Is there any potentially malicious behavior? Briefly answer `Yes' or `No'. If yes, explain why.}'' to ask GPT-4 to perform classification of malicious and non-malicious code.

In both cases, we observe that GPT-4 is able to correctly explain each function. Additionally, it successfully points out malicious activities inside the code of \texttt{KratosKnife}. Based on the generated analysis, GPT-4 correctly classifies code files that contain malicious behaviors. There are some interesting observations in this experiment:

\begin{description}
    \item[Observation 1:] GPT-4 is able to point out malicious behaviors by recognizing typical patterns.
\end{description}
For example, when analyzing the code of malware \texttt{KratosKnife}, GPT-4 is able to point out that the VM environment checking is malicious since its purpose is to counter malware analyses. 

\begin{description}
    \item[Observation 2:] There are potential false alarms in the classification step.
\end{description}
In the codebase of \texttt{KratosKnife}, there exist cryptography utility functions designed to encrypt code files based on an input key provided to the Python script. This encryption process serves to obfuscate code. Notably, these utility functions operate independently of other malicious behaviors associated with the malware. Furthermore, the analysis generated does not indicate any inherently malicious behaviors. However, during the classification process, GPT-4 considers these utility functions malicious since these functions encrypt files, and falsely accuse them of replacing encrypted files on disk, even though the generated explanation indicates the encrypted code is stored in separate files. This discrepancy underscores the necessity for a thorough examination of classification results produced by LLMs.

\subsection{Mobile Platform Virus Analysis}
\begin{figure}[htbp]
    \centering
    \includegraphics[width=.7\linewidth]{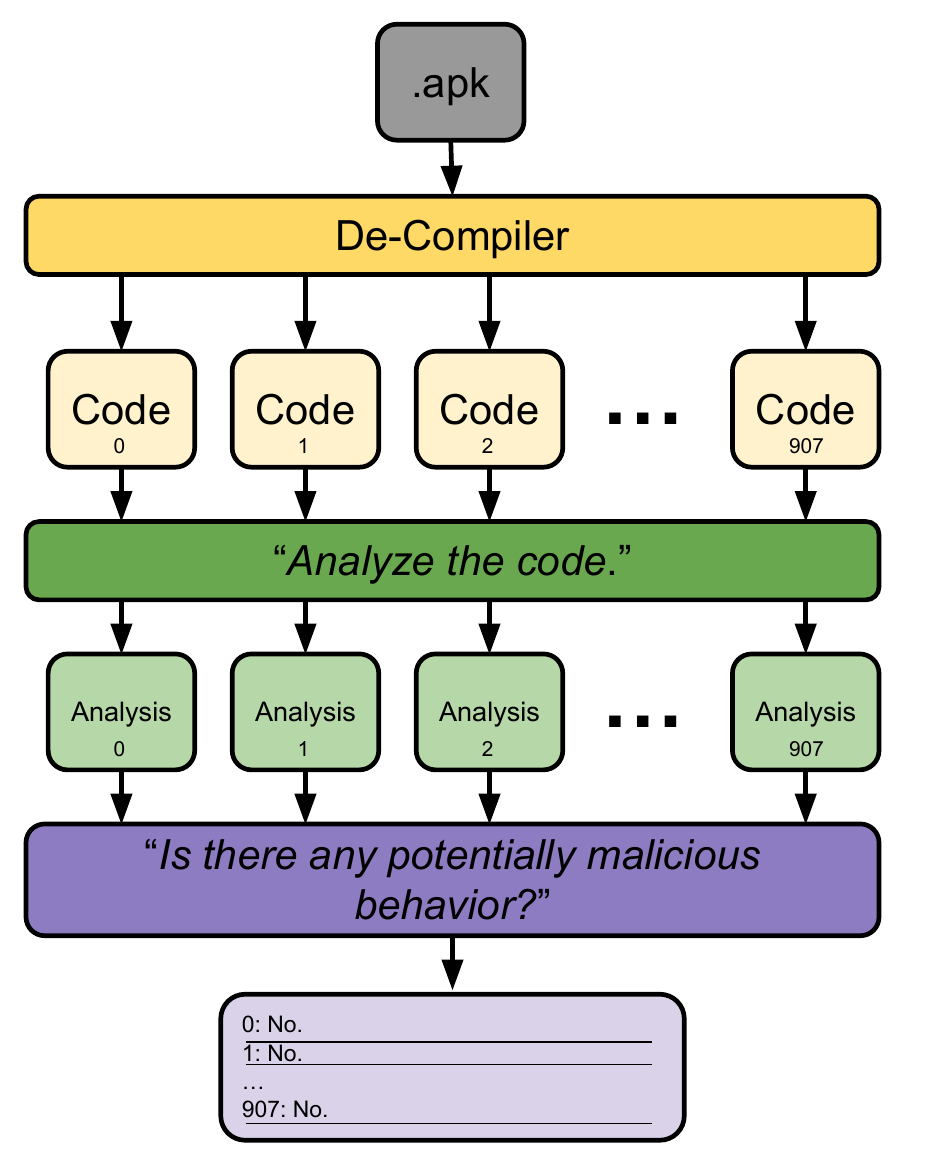}
    \caption{Diagram of our Android case study pipeline.}
    \label{FigAndroidDiagram}
\end{figure}

In our first case study for decompiled code, we focus on analyzing a decompiled Android virus from the Internet. We utilize 
Java Decompiler~\cite{javadecompiler}
to decompile an android virus called \texttt{msg-stealer}, obtained from \cite{androsmsstealer}. As indicated by its name, this virus maliciously steals SMS messages from Android mobile phone users. By using 
\texttt{jd-gui}, a graphical utility provided by the Java Decompiler
, we successfully decompile the \texttt{.apk} file of the virus and generate $908$ \texttt{.java} files, containing $\sim 250$k LoC. The experiment process is identical to our previous case study: we feed every decompiled code file to both GPT-3.5 and GPT-4 and get $908$ code analysis responses in return. After that, we instruct GPT-4 to read the analysis and give alerts if there are any abnormal or potentially malicious behaviors. The diagram of our experiment pipeline is shown in Figure~\ref{FigAndroidDiagram}.

This virus works in the following way:
\begin{enumerate}
    \item Upon signing up, it requires users to enter a phone number. It checks if the entered phone number is from Iran (+98). If so, it requests to gain SMS sending and reading privileges.
    \item It then continues to establish a connection using a specific URL.
    \item When an SMS message is received, it reads the whole packet of information and performs a string match. Upon a successful match, the virus sends the content of the SMS message to a specific site.
\end{enumerate}
These functionalities are spread across $3$ different code files.

By examining the analysis generated by GPT-3.5 and GPT-4, we find that both have correctly deciphered these behaviors from the decompiled source code. Part of the responses from GPT-3.5 are shown below:
\begin{description}
    \item[Response 1:] \textit{``... the code validates the entered phone number using a regular expression. If the phone number is not valid, a toast message is displayed. Otherwise, the code requests the `RECEIVE\_SMS' permission and checks if the permission is granted.''} (Step 1)
    \item[Response 2:] \textit{``... it builds a URL string by appending the `url' and `info'} (user phone number) \textit{parameters to a base URL. It then sends a GET request to this URL using the AndroidNetworking library's `get()` method.''} (Step 2)
    \item[Response 3:] \textit{``... it retrieves the SMS messages from the intent extras, retrieves the message body, and concatenates all message bodies into a single string. It then checks if the string contains the text "}(a trigger string in Persian)\textit{" and if so, it updates a shared preference value named "lock" to "off". Finally, it uses the `connect` class to perform some action using the phone number retrieved from the shared preferences and the concatenated string of SMS messages.''} (Step 3)
\end{description}

These behaviors are apparently suspicious, since it requests SMS read/write privilege (sensitive privilege) and communicates with and sends SMS messages to a remote site. However, when we feed GPT-4/GPT-3.5 with its analysis results and ask \textit{``Is there any potentially malicious behavior?''}, we have the following interesting observations.
\begin{description}
    \item[Observation 1:] When feeding analysis results regarding the three steps independently, both GPT-3.5 and GPT-4 fail to recognize the potentially malicious behaviors.
    \item[Observation 2:] When concatenating the information together and feeding to LLMs, only GPT-4 successfully points out potentially malicious behaviors.
\end{description}

These observations indicate that GPT-4 possesses a better capability to understand natural language inputs and GPT-4 would be a better choice to process the massive analysis results. However, our observations also suggest that when using these LLMs for defensive analysis, it is important to provide enough context otherwise LLMs will fail to detect security issues.

\subsection{WannaCry Ransomware Analysis}
In another case study, we aim for a more realistic target and analyze a desktop virus to showcase if LLMs can be utilized to perform defensive analysis on decompiled code. We select a malicious software called WannaCry (also named WanaCrypt0r)~\cite{mohurle2017brief}, which is a ransomware that posted shockingly devastating damages to various organizations and personal users in 2017. WannaCry ransomware attack uses the EternalBlue exploit to compromise Windows Service Message Block Protocol (SMBv1) and propagates itself through TCP port 445~\cite{national2017wannacry,kao2018dynamic}. Once residing on a computer, it encrypts user files and demands a ransom paid in bitcoins. It is reported that WannaCry ransomware has impacted hospitals~\cite{hospitalnews} and large companies including Nissan~\cite{nissannews}, TSMC~\cite{tsmcnews}, etc.

In this case study, we directly feed the decompiled code file of WannaCry~\cite{wannacrycode} generated by an open-source decompiler RetDec~\cite{RetDec} to LLMs. We aim to analyze the behaviors of WannaCry, and test the ability of LLMs to understand and analyze code. The decompiled code file contains around $11$k lines of C code and automatically generated comments. The decompiled C code does not include meaningful functions and variable names. It has identifier names such as \texttt{function\_401000} and \texttt{g33} that do not provide useful contextual information. The automatically generated comments only specify memory addresses of functions and branches, as well as a list of functions that are statically or dynamically linked.

To feed the generated code to LLMs, we input the generated code to LLMs in an interactive manner. In the first prompt, we state that we will input a code file for it to analyze, and it may consist of multiple chunks of code due to the length restrictions of input. For shorter functions, we copy and paste $3$-$5$ functions in one round of query and obtain the response of analysis. For longer functions whose length exceeds the length limit of input, we partition the body of functions into multiple parts, each consisting of $200$-$500$ lines of code. A sentence is first provided to LLMs, stating that this is a long function and will be input as multiple chunks of code. Starting from the second chunk of code, we state that ``\textit{the following code snippet is the remaining part of the previous function}''. After the last partition of code, we manually add a comment ``\texttt{// function ends}'' to indicate the end of the function. No other manual operation is involved during the experiment. We select to use GPT-3.5 in this experiment, since it is one of the easiest accessible and most popular chatbot-based LLMs. 

Part of the analysis results from GPT-3.5 that we consider interesting and important are shown below. For the whole record of our experiment, please refer to our online appendix.

\begin{description}
\item[Response 1:] \textit{``... Overall, these functions handle networking tasks. The first, \texttt{function\_4072a0}, seems to be part of a more complex communication protocol, while the second, \texttt{function\_407480}, checks for the ability to connect to port 445.''}
\item[Response 2:] \textit{``... This is interesting functionality, especially the last function. The program is attempting to create a Windows service named ``\texttt{mssecsvc2.0}'', which sounds similar to a legitimate Microsoft Security service. This could indicate malicious or at least suspicious behavior, as it seems like the program is trying to disguise itself as a legitimate service.''} (\texttt{function\_407c40})
\item[Response 3:] \textit{``...$\bullet$ Initializes a URL string pointing to http://www.iuqerfsodp9ifjap\{...\}jfaewrwergwea.com.\\
    $\bullet$ Uses the Windows Internet (WinINet) API to open an internet session and attempts to open the URL.\\
    $\bullet$ Closes the internet handles after the operations.\\
    $\bullet$ If it's successful in opening the URL, it closes the internet handle and returns, else it calls \texttt{function\_408090}.\\
    $\bullet$ This function seems to be checking for the availability of a specific website, possibly as a kind of ``alive'' check or to determine network connectivity. ...''} (\texttt{function\_408140})
\item[Response 4:] \textit{``... The two goto labels \texttt{lab\_0x408755\_9} and \texttt{lab\_0x4087c3\_9} are exit points or continuation points, but the actual code for these labels is not present in the segment. They might direct the code to other parts of the function based on specific conditions. ...''}
\item[Response 5:] \textit{``... From the provided data, it's challenging to pinpoint the exact purpose of the application, but it seems like a complex program that could serve as a service or a daemon on Windows, handling network communication, possibly acting as a server or a service manager, with capabilities for file handling and other utility tasks. However, this is a speculative overview based on the given context, and a more detailed analysis would be required to confirm this.''}
\end{description}

From the findings presented above, we found that GPT-3.5 is capable of analyzing a relatively large-scale decompiled c codebase with low readability. It correctly extracts critical behaviors of our target ransomware, including:
\begin{itemize}
    \item Checking connectivity to port 445 (Response 1);
    \item Creating service named \texttt{mssecsvc2.0} (Response 2);
    \item Querying a special url for ``alive'' check (Response 3).
\end{itemize}
These critical findings partially align with publicly available analysis of WannaCry~\cite{microsoftblogwannacry}, as these are also considered important behavioral watermarks of WannaCry ransomware. However, there are differences indicating that the analysis is not fully correct. For example, according to the technical report provided by Microsoft~\cite{microsoftblogwannacry}, the purpose of initiating \texttt{mssecsvc2.0} service is to exploit the SMB vulnerability, instead of simply disguising itself. However, considering the lack of context information and the fact that GPT-3.5 correctly raises security concerns, this still demonstrates the capability of LLMs in defensive analysis.

However, there are other observed weaknesses of GPT-3.5:
\begin{description}
    \item[Observation 1:] Code address labels are ignored and not remembered.
\end{description}
In Response 4, GPT-3.5 replies that code for two labels is not provided. Yet the code labels actually reside in the same function body and were provided to GPT-3.5 a few rounds of queries ago due to the excessively long function body. By looking back at the dialog record, we find that these labels are also not mentioned in the response at the corresponding round of query. This indicates that these labels are ignored, and not remembered during later processing. The incapability to capture and remember important code information will hinder the ability of LLMs to analyze long code bodies with more complicated context, considering the rather strict input limit of each query.

\begin{description}
    \item[Observation 2:] GPT-3.5 is conservative in providing conclusive remarks.
\end{description}
During our experiment, GPT-3.5 repeatedly mentions the need for more context and its incapability to draw a conclusion about the intent of the provided code. In the conclusive Response 5, it simply generates generic descriptions. We suspect that GPT-3.5 learned the strategy to conservatively provide information to improve the score of outputs during training.

\noindent
\textbf{Summary.}
In this case study, the analysis results of WannaCry show that GPT-3.5 is able to capture the most important behaviors of this ransomware and successfully provide a security alert (Response 2). Despite the observed weaknesses, we confirm that LLM models like GPT-3.5 are able to extract important information from decompiled code and raise security concerns if there are any. This experiment shows LLMs' superior ability in performing code analysis and summarization, considering that the provided code has extremely low readability and there is no extra context. Though at this point, LLMs cannot replace expert human beings, they can still serve as an initial code analysis assistant to accelerate the reverse engineering of software and aid defensive code analysis.

\section{Discussion}\label{SecDiss}

\subsection{Using LLMs for Code Analysis}
As shown in our analysis results, more advanced LLMs (e.g. GPT-3.5, GPT-4) have a higher success rate of generating explanations for input code samples. However, as found out in our analysis, smaller ones have poor performance, and the results are not always reliable even for larger LLMs, especially when complicated code obfuscation is involved. Smaller general-purpose LLMs, even when specifically fine-tuned for coding tasks, struggle to produce meaningful results. These findings suggest that at this point, utilizing larger LLMs is still the safer choice for code analysis-related tasks.

\subsection{Future Work}
This work sheds light on some possible future directions in this area. First of all, there is a gap in the literature about obfuscated code datasets. Most works only focus on the ability to analyze normal code, without integrating obfuscation techniques. Constructing such a large dataset would enable LLM users to fine-tune pre-trained LLMs specifically for code analysis tasks quickly. Second, in this paper, we report several interesting observations regarding the memorization phenomena of LLMs~\cite{carlini2021extracting}. The underlying mechanisms still remain to be explored. Third, regarding similarity metrics, while the n-gram algorithm-based metric has been reported to be suboptimal in capturing semantic similarities~\cite{yuan2023evaluating}, during our experiments, the newly employed ChatGPT-based method~\cite{zheng2023judging,chiang2023can,yuan2023evaluating,bubeck2023sparks} is also not reliable and requires manual work to calibrate, especially when code snippets and natural languages are combined together in the inputs. We believe constructing a more sophisticated metric that captures the essence of code analysis results is non-trivial and could be one research direction.

\section{Related Work}\label{SecRelWork}
\noindent
\textbf{Code Analysis.}
Code analysis is crucial in modern software development to identify vulnerabilities and ensure code quality. 
The significant progress in artificial intelligence (AI) motivates researchers to adopt advanced AI models to improve more effective code analysis. Mou \etal~\cite{mou2016convolutional} proposed to use CNNs to analyze syntax trees of programs. Feng \etal~\cite{feng2020efficient} targets detecting vulnerability by collecting features from extracted syntax trees and employing a Bi-GRU network to identify bugs in general code. 
The development of natural language processing techniques, especially transformers and LLMs, significantly boost this area. Chen \etal~\cite{chen2022augmenting} employ a transformer-based method to automatically predict variable names and types from decompiler outputs, thus dramatically improving the readability of decompiler-generated code. Similarly, Xu \etal~\cite{xu2023lmpa} also proposes to use LLMs to assist in handling decompilation tasks. It presents that using an iterative algorithm with multiple LLM queries can improve the decompilation results.

\noindent
\textbf{LLM Evaluation.}
As LLMs become prevalent in recent years, there are a surging number of review or analysis papers focusing on LLM's capabilities. Bubeck \etal~\cite{bubeck2023sparks} analyze the capabilities of GPT-4 from different aspects, including vision, coding, and mathematics-related tasks and discuss the societal impacts of such LLMs. A recent survey~\cite{barrett2023identifying} focuses on security aspects of LLMs and summarizes several attack and defence works on LLMs and points out future directions of research in this field. Regarding analysis and evaluation of code-related capabilities of LLMs, Chen \etal~\cite{chen2021evaluating} evaluate LLMs trained on code. However, their focus is primarily on the code generation aspects. The work that is closest to ours is~\cite{yuan2023evaluating}. In the paper, they conduct an analysis of the ability of instruction-tuned and fine-tuned LLMs to perform code comprehension and generation tasks. However, code-obfuscation-related tasks, including obfuscated code comprehension and de-obfuscated code generation, are not included.
\section{Conclusion}\label{SecConc}
In this paper, we have conducted a thorough evaluation of the code analysis capabilities of popular Large Language Models (LLMs). Our results reveal that the larger LLMs, specifically from the GPT series, exhibit impressive performance in code analysis tasks. On the other hand, the smaller models from LLaMA family evaluated in this paper demonstrate unsatisfying performance in these same tasks. When it comes to analyzing obfuscated code, the GPT-series LLMs still produce reasonably useful results for explanation-related tasks; however, they do encounter limitations in providing de-obfuscated code. Our study also uncovers intriguing findings such as LLM memorization phenomena. Our research highlights that, at the present stage, LLMs demonstrate considerable potential in this field. However, there are still many unexplored ways to optimize their performance.

\section*{Acknowledgments}
We would like to thank the authors of Wobfuscator~\cite{romano2022wobfuscator} for sharing their obfuscation tool. We extend our gratitude to all anonymous reviewers and our shepherd for the valuable feedback they have provided.




\bibliographystyle{plain}
\bibliography{mybib}

\begin{thebibliography}{10}

\bibitem{javadecompiler}
Java decompiler.
\newblock \url{http://java-decompiler.github.io/}.

\bibitem{kratosknife}
Kratosknife.
\newblock \url{https://github.com/PushpenderIndia/KratosKnife}.

\bibitem{librarian}
librarian.
\newblock \url{https://github.com/oto-labs/librarian/tree/main}.

\bibitem{androsmsstealer}
Not so boring android malware.
\newblock \url{https://maldroid.github.io/android-malware-samples//}.

\bibitem{tripforce}
tripforce.
\newblock \url{https://github.com/microsounds/tripforce}, 2016.

\bibitem{semantictextsimilarity}
semantic-text-similarity.
\newblock \url{https://github.com/AndriyMulyar/semantic-text-similarity}, 2019.

\bibitem{jsapps}
js-apps.
\newblock \url{https://github.com/jaiimeriios/js-apps}, 2022.

\bibitem{wannacrycode}
Wannacry/worm.c.
\newblock \url{https://github.com/xcp3r/WannaCry/blob/main/worm.c}, 2022.

\bibitem{poj104}
Codexglue/code-code/clone-detection-poj-104/.
\newblock \url{https://github.com/microsoft/CodeXGLUE/blob/main/Code-Code/Clone-detection-POJ-104/README.md}, 2023.

\bibitem{heron}
incubator/heron.
\newblock \url{https://github.com/apache/incubator-heron}, 2023.

\bibitem{ioccc}
The international obfuscated c code contest.
\newblock \url{https://www.ioccc.org/}, 2023.

\bibitem{jsobtool}
Javascript obfuscator tool.
\newblock \url{https://obfuscator.io/}, 2023.

\bibitem{octane}
Octane 2.0.
\newblock \url{https://chromium.github.io/octane/}, 2023.

\bibitem{RetDec}
Retdec.
\newblock \url{https://github.com/avast/retdec}, 2023.

\bibitem{acm}
ACM.
\newblock Acm policy on authorship.
\newblock \url{https://www.acm.org/publications/policies/new-acm-policy-on-authorship}, 2023.

\bibitem{Ahmed2023-FewshotTrainingLLMs}
Toufique Ahmed and Premkumar Devanbu.
\newblock Few-shot training {{LLMs}} for project-specific code-summarization.
\newblock In {\em Proceedings of the 37th {{IEEE}}/{{ACM International Conference}} on {{Automated Software Engineering}}}, {{ASE}} '22, pages 1--5, {New York, NY, USA}, January 2023. {Association for Computing Machinery}.

\bibitem{araci2019finbert}
Dogu Araci.
\newblock Finbert: Financial sentiment analysis with pre-trained language models.
\newblock {\em arXiv preprint arXiv:1908.10063}, 2019.

\bibitem{Bairi2023-CodePlanRepositorylevelCoding}
Ramakrishna Bairi, Atharv Sonwane, Aditya Kanade, Vageesh~D. C, Arun Iyer, Suresh Parthasarathy, Sriram Rajamani, B.~Ashok, and Shashank Shet.
\newblock {{CodePlan}}: {{Repository-level Coding}} using {{LLMs}} and {{Planning}}, September 2023.

\bibitem{Barke2023-GroundedCopilotHow}
Shraddha Barke, Michael~B. James, and Nadia Polikarpova.
\newblock Grounded {{Copilot}}: {{How Programmers Interact}} with {{Code-Generating Models}}.
\newblock {\em Proceedings of the ACM on Programming Languages}, 7(OOPSLA1):78:85--78:111, April 2023.

\bibitem{barrett2023identifying}
Clark Barrett, Brad Boyd, Ellie Burzstein, Nicholas Carlini, Brad Chen, Jihye Choi, Amrita~Roy Chowdhury, Mihai Christodorescu, Anupam Datta, Soheil Feizi, et~al.
\newblock Identifying and mitigating the security risks of generative ai.
\newblock {\em arXiv preprint arXiv:2308.14840}, 2023.

\bibitem{binkley2007source}
David Binkley.
\newblock Source code analysis: A road map.
\newblock {\em Future of Software Engineering (FOSE'07)}, pages 104--119, 2007.

\bibitem{Bird2023-TakingFlightCopilot}
Christian Bird, Denae Ford, Thomas Zimmermann, Nicole Forsgren, Eirini Kalliamvakou, Travis Lowdermilk, and Idan Gazit.
\newblock Taking {{Flight}} with {{Copilot}}.
\newblock {\em Communications of the ACM}, 66(6):56--62, May 2023.

\bibitem{brown1992class}
Peter~F Brown, Vincent~J Della~Pietra, Peter~V Desouza, Jennifer~C Lai, and Robert~L Mercer.
\newblock Class-based n-gram models of natural language.
\newblock {\em Computational linguistics}, 18(4):467--480, 1992.

\bibitem{bubeck2023sparks}
S{\'e}bastien Bubeck, Varun Chandrasekaran, Ronen Eldan, Johannes Gehrke, Eric Horvitz, Ece Kamar, Peter Lee, Yin~Tat Lee, Yuanzhi Li, Scott Lundberg, et~al.
\newblock Sparks of artificial general intelligence: Early experiments with gpt-4.
\newblock {\em arXiv preprint arXiv:2303.12712}, 2023.

\bibitem{canfora2011achievements}
Gerardo Canfora, Massimiliano Di~Penta, and Luigi Cerulo.
\newblock Achievements and challenges in software reverse engineering.
\newblock {\em Communications of the ACM}, 54(4):142--151, 2011.

\bibitem{cao2021bgnn4vd}
Sicong Cao, Xiaobing Sun, Lili Bo, Ying Wei, and Bin Li.
\newblock Bgnn4vd: Constructing bidirectional graph neural-network for vulnerability detection.
\newblock {\em Information and Software Technology}, 136:106576, 2021.

\bibitem{carlini2021extracting}
Nicholas Carlini, Florian Tramer, Eric Wallace, Matthew Jagielski, Ariel Herbert-Voss, Katherine Lee, Adam Roberts, Tom Brown, Dawn Song, Ulfar Erlingsson, et~al.
\newblock Extracting training data from large language models.
\newblock In {\em 30th USENIX Security Symposium (USENIX Security 21)}, pages 2633--2650, 2021.

\bibitem{cascella2023evaluating}
Marco Cascella, Jonathan Montomoli, Valentina Bellini, and Elena Bignami.
\newblock Evaluating the feasibility of chatgpt in healthcare: an analysis of multiple clinical and research scenarios.
\newblock {\em Journal of Medical Systems}, 47(1):33, 2023.

\bibitem{chen2021evaluating}
Mark Chen, Jerry Tworek, Heewoo Jun, Qiming Yuan, Henrique Ponde de~Oliveira Pinto, Jared Kaplan, Harri Edwards, Yuri Burda, Nicholas Joseph, Greg Brockman, et~al.
\newblock Evaluating large language models trained on code.
\newblock {\em arXiv preprint arXiv:2107.03374}, 2021.

\bibitem{chen2022augmenting}
Qibin Chen, Jeremy Lacomis, Edward~J Schwartz, Claire Le~Goues, Graham Neubig, and Bogdan Vasilescu.
\newblock Augmenting decompiler output with learned variable names and types.
\newblock In {\em 31st USENIX Security Symposium (USENIX Security 22)}, pages 4327--4343, 2022.

\bibitem{chiang2023can}
Cheng-Han Chiang and Hung-yi Lee.
\newblock Can large language models be an alternative to human evaluations?
\newblock {\em arXiv preprint arXiv:2305.01937}, 2023.

\bibitem{christodorescu2003static}
Mihai Christodorescu and Somesh Jha.
\newblock Static analysis of executables to detect malicious patterns.
\newblock In {\em 12th USENIX Security Symposium (USENIX Security 03)}, 2003.

\bibitem{nissannews}
ChronicleLive.
\newblock Ransomware cyber attack recap: Nissan confirm they have been hit by hack which crippled {NHS}.
\newblock \url{https://www.chroniclelive.co.uk/news/north-east-news/cyber-attack-nhs-latest-news-13029913}, May 2017.

\bibitem{curnow1976synthetic}
Harold~J Curnow and Brian~A. Wichmann.
\newblock A synthetic benchmark.
\newblock {\em The Computer Journal}, 19(1):43--49, 1976.

\bibitem{dongarra2003linpack}
Jack~J Dongarra, Piotr Luszczek, and Antoine Petitet.
\newblock The linpack benchmark: past, present and future.
\newblock {\em Concurrency and Computation: practice and experience}, 15(9):803--820, 2003.

\bibitem{coremark}
{EEMBC}.
\newblock Coremark.
\newblock \url{https://www.eembc.org/coremark/}, 2022.

\bibitem{eidreverse}
Abdelrahman Eid.
\newblock Reverse engineering snapchat (part i): Obfuscation techniques.
\newblock \url{https://hot3eed.github.io/snap_part1_obfuscations.html}.

\bibitem{fang2022towards}
Ruijie Fang, Ruoyu Zhang, Elahe Hosseini, Anna~M Parenteau, Sally Hang, Setareh Rafatirad, Camelia~E Hostinar, Mahdi Orooji, and Houman Homayoun.
\newblock Towards generalized ml model in automated physiological arousal computing: A transfer learning-based domain generalization approach.
\newblock In {\em 2022 IEEE International Conference on Bioinformatics and Biomedicine (BIBM)}, pages 2577--2584. IEEE, 2022.

\bibitem{feng2020efficient}
Hantao Feng, Xiaotong Fu, Hongyu Sun, He~Wang, and Yuqing Zhang.
\newblock Efficient vulnerability detection based on abstract syntax tree and deep learning.
\newblock In {\em IEEE INFOCOM 2020-IEEE Conference on Computer Communications Workshops (INFOCOM WKSHPS)}, pages 722--727. IEEE, 2020.

\bibitem{githubtop}
Github.
\newblock The top programming languages.
\newblock \url{https://octoverse.github.com/2022/top-programming-languages}, 2022.

\bibitem{copilot}
Github.
\newblock Github copilot.
\newblock \url{https://github.com/features/copilot}, 2023.

\bibitem{gustafson1995hint}
John~L Gustafson and Quinn~O Snell.
\newblock Hint: A new way to measure computer performance.
\newblock In {\em Proceedings of the Twenty-Eighth Annual Hawaii International Conference on System Sciences}, volume~2, pages 392--401. IEEE, 1995.

\bibitem{haas2017bringing}
Andreas Haas, Andreas Rossberg, Derek~L Schuff, Ben~L Titzer, Michael Holman, Dan Gohman, Luke Wagner, Alon Zakai, and JF~Bastien.
\newblock Bringing the web up to speed with webassembly.
\newblock In {\em Proceedings of the 38th ACM SIGPLAN Conference on Programming Language Design and Implementation}, pages 185--200, 2017.

\bibitem{he2016deep}
Kaiming He, Xiangyu Zhang, Shaoqing Ren, and Jian Sun.
\newblock Deep residual learning for image recognition.
\newblock In {\em Proceedings of the IEEE conference on computer vision and pattern recognition}, pages 770--778, 2016.

\bibitem{hosseini2022convolution}
Elahe Hosseini, Ruijie Fang, Ruoyu Zhang, Chen-Nee Chuah, Mahdi Orooji, Soheil Rafatirad, Setareh Rafatirad, and Houman Homayoun.
\newblock Convolution neural network for pain intensity assessment from facial expression.
\newblock In {\em 2022 44th annual international conference of the IEEE Engineering in Medicine \& Biology Society (EMBC)}, pages 2697--2702. IEEE, 2022.

\bibitem{husain2019codesearchnet}
Hamel Husain, Ho-Hsiang Wu, Tiferet Gazit, Miltiadis Allamanis, and Marc Brockschmidt.
\newblock Codesearchnet challenge: Evaluating the state of semantic code search.
\newblock {\em arXiv preprint arXiv:1909.09436}, 2019.

\bibitem{ieee}
IEEE.
\newblock Submission and peer review policies.
\newblock \url{https://journals.ieeeauthorcenter.ieee.org/become-an-ieee-journal-author/publishing-ethics/guidelines-and-policies/submission-and-peer-review-policies/}, 2023.

\bibitem{jessen2023chit}
Urszula Jessen, Michal Sroka, and Dirk Fahland.
\newblock Chit-chat or deep talk: {{Prompt}} engineering for process mining.
\newblock {\em arXiv preprint arXiv:2307.09909}, 2023.

\bibitem{kao2018dynamic}
Da-Yu Kao and Shou-Ching Hsiao.
\newblock The dynamic analysis of wannacry ransomware.
\newblock In {\em 2018 20th International conference on advanced communication technology (ICACT)}, pages 159--166. IEEE, 2018.

\bibitem{kasneci2023chatgpt}
Enkelejda Kasneci, Kathrin Se{\ss}ler, Stefan K{\"u}chemann, Maria Bannert, Daryna Dementieva, Frank Fischer, Urs Gasser, Georg Groh, Stephan G{\"u}nnemann, Eyke H{\"u}llermeier, et~al.
\newblock Chatgpt for good? on opportunities and challenges of large language models for education.
\newblock {\em Learning and individual differences}, 103:102274, 2023.

\bibitem{laszlo2009obfuscating}
T{\i}mea L{\'a}szl{\'o} and {\'A}kos Kiss.
\newblock Obfuscating c++ programs via control flow flattening.
\newblock {\em Annales Universitatis Scientarum Budapestinensis de Rolando E{\"o}tv{\"o}s Nominatae, Sectio Computatorica}, 30(1):3--19, 2009.

\bibitem{Leinonen2023-ComparingCodeExplanations}
Juho Leinonen, Paul Denny, Stephen MacNeil, Sami Sarsa, Seth Bernstein, Joanne Kim, Andrew Tran, and Arto Hellas.
\newblock Comparing {{Code Explanations Created}} by {{Students}} and {{Large Language Models}}, April 2023.

\bibitem{Li2024-EnhancingLLMBasedCoding}
Yichen Li, Yun Peng, Yintong Huo, and Michael~R. Lyu.
\newblock Enhancing {{LLM-Based Coding Tools}} through {{Native Integration}} of {{IDE-Derived Static Context}}, February 2024.

\bibitem{liu2021software}
Binbin Liu, Weijie Feng, Qilong Zheng, Jing Li, and Dongpeng Xu.
\newblock Software obfuscation with non-linear mixed boolean-arithmetic expressions.
\newblock In {\em Information and Communications Security: 23rd International Conference, ICICS 2021, Chongqing, China, November 19-21, 2021, Proceedings, Part I 23}, pages 276--292. Springer, 2021.

\bibitem{livshits2010zozzle}
Charles Curtsinger~Benjamin Livshits, Ben Zorn, and Christian Seifert.
\newblock Zozzle: Low-overhead mostly static javascript malware detection.
\newblock In {\em USENIX Security Symposium}, 2010.

\bibitem{mccalpin1995memory}
John~D McCalpin et~al.
\newblock Memory bandwidth and machine balance in current high performance computers.
\newblock {\em IEEE computer society technical committee on computer architecture (TCCA) newsletter}, 2(19-25), 1995.

\bibitem{microsoftblogwannacry}
Microsoft.
\newblock Wannacrypt ransomware worm targets out-of-date systems.
\newblock \url{https://www.microsoft.com/en-us/security/blog/2017/05/12/wannacrypt-ransomware-worm-targets-out-of-date-systems/}, May 2017.

\bibitem{mohurle2017brief}
Savita Mohurle and Manisha Patil.
\newblock A brief study of wannacry threat: Ransomware attack 2017.
\newblock {\em International journal of advanced research in computer science}, 8(5):1938--1940, 2017.

\bibitem{mou2016convolutional}
Lili Mou, Ge~Li, Lu~Zhang, Tao Wang, and Zhi Jin.
\newblock Convolutional neural networks over tree structures for programming language processing.
\newblock In {\em Proceedings of the AAAI conference on artificial intelligence}, volume~30, 2016.

\bibitem{national2017wannacry}
{National Cybersecurity and Communications Integration Center}.
\newblock What is wannacry/wanacrypt0r?, 2017.

\bibitem{neamtiu2005understanding}
Iulian Neamtiu, Jeffrey~S Foster, and Michael Hicks.
\newblock Understanding source code evolution using abstract syntax tree matching.
\newblock In {\em Proceedings of the 2005 international workshop on Mining software repositories}, pages 1--5, 2005.

\bibitem{hospitalnews}
Sky News.
\newblock {NHS} cyberattack: List of hospitals hit by ransomware strike.
\newblock \url{https://news.sky.com/story/nhs-cyberattack-full-list-of-organisations-affected-so-far-10874493}, May 2017.

\bibitem{tsmcnews}
The~Hawker News.
\newblock Tsmc chip maker blames wannacry malware for production halt.
\newblock \url{https://thehackernews.com/2018/08/tsmc-wannacry-ransomware-attack.html}, August 2018.

\bibitem{chatgpt}
OpenAI.
\newblock Chatgpt.
\newblock \url{https://chat.openai.com/}, 2023.

\bibitem{seehearspeak}
OpenAI.
\newblock Chatgpt can now see, hear, and speak.
\newblock \url{https://openai.com/blog/chatgpt-can-now-see-hear-and-speak}, 2023.

\bibitem{openai2023gpt4}
OpenAI.
\newblock Gpt-4 technical report.
\newblock 2023.

\bibitem{nbench}
PetaBridge.
\newblock {NBench}.
\newblock \url{https://nbench.io/}, 2023.

\bibitem{Prather2023-RobotsAreHere}
James Prather, Paul Denny, Juho Leinonen, Brett~A. Becker, Ibrahim Albluwi, Michelle Craig, Hieke Keuning, Natalie Kiesler, Tobias Kohn, Andrew {Luxton-Reilly}, Stephen MacNeil, Andrew Peterson, Raymond Pettit, Brent~N. Reeves, and Jaromir Savelka.
\newblock The {{Robots}} are {{Here}}: {{Navigating}} the {{Generative AI Revolution}} in {{Computing Education}}.
\newblock In {\em Proceedings of the 2023 {{Working Group Reports}} on {{Innovation}} and {{Technology}} in {{Computer Science Education}}}, pages 108--159, December 2023.

\bibitem{Prather2024-InteractionsPromptProblems}
James Prather, Paul Denny, Juho Leinonen, David~H. Smith~IV, Brent~N. Reeves, Stephen MacNeil, Brett~A. Becker, Andrew {Luxton-Reilly}, Thezyrie Amarouche, and Bailey Kimmel.
\newblock Interactions with {{Prompt Problems}}: {{A New Way}} to {{Teach Programming}} with {{Large Language Models}}, January 2024.

\bibitem{romano2022wobfuscator}
Alan Romano, Daniel Lehmann, Michael Pradel, and Weihang Wang.
\newblock Wobfuscator: Obfuscating javascript malware via opportunistic translation to webassembly.
\newblock In {\em 2022 IEEE Symposium on Security and Privacy (SP)}, pages 1574--1589. IEEE, 2022.

\bibitem{roziere2023code}
Baptiste Rozi{\`e}re, Jonas Gehring, Fabian Gloeckle, Sten Sootla, Itai Gat, Xiaoqing~Ellen Tan, Yossi Adi, Jingyu Liu, Tal Remez, J{\'e}r{\'e}my Rapin, et~al.
\newblock Code llama: Open foundation models for code.
\newblock {\em arXiv preprint arXiv:2308.12950}, 2023.

\bibitem{salman2015students}
Iflaah Salman, Ayse~Tosun Misirli, and Natalia Juristo.
\newblock Are students representatives of professionals in software engineering experiments?
\newblock In {\em 2015 IEEE/ACM 37th IEEE international conference on software engineering}, volume~1, pages 666--676. IEEE, 2015.

\bibitem{sanderson2023gpt}
Katharine Sanderson.
\newblock Gpt-4 is here: what scientists think.
\newblock {\em Nature}, 615(7954):773, 2023.

\bibitem{solaiman2019release}
Irene Solaiman, Miles Brundage, Jack Clark, Amanda Askell, Ariel Herbert-Voss, Jeff Wu, Alec Radford, Gretchen Krueger, Jong~Wook Kim, Sarah Kreps, et~al.
\newblock Release strategies and the social impacts of language models.
\newblock {\em arXiv preprint arXiv:1908.09203}, 2019.

\bibitem{taori2023alpaca}
Rohan Taori, Ishaan Gulrajani, Tianyi Zhang, Yann Dubois, Xuechen Li, Carlos Guestrin, Percy Liang, and Tatsunori~B Hashimoto.
\newblock Alpaca: A strong, replicable instruction-following model.
\newblock {\em Stanford Center for Research on Foundation Models. https://crfm. stanford. edu/2023/03/13/alpaca. html}, 3(6):7, 2023.

\bibitem{topol2019high}
Eric~J Topol.
\newblock High-performance medicine: the convergence of human and artificial intelligence.
\newblock {\em Nature medicine}, 25(1):44--56, 2019.

\bibitem{touvron2023llama}
Hugo Touvron, Thibaut Lavril, Gautier Izacard, Xavier Martinet, Marie-Anne Lachaux, Timoth{\'e}e Lacroix, Baptiste Rozi{\`e}re, Naman Goyal, Eric Hambro, Faisal Azhar, et~al.
\newblock Llama: Open and efficient foundation language models.
\newblock {\em arXiv preprint arXiv:2302.13971}, 2023.

\bibitem{Tunstall2023starchat-alpha}
Lewis Tunstall, Nathan Lambert, Nazneen Rajani, Edward Beeching, Teven Le~Scao, Leandro von Werra, Sheon Han, Philipp Schmid, and Alexander Rush.
\newblock Creating a coding assistant with starcoder.
\newblock {\em Hugging Face Blog}, 2023.
\newblock https://huggingface.co/blog/starchat.

\bibitem{vaithilingam2022expectation}
Priyan Vaithilingam, Tianyi Zhang, and Elena~L Glassman.
\newblock Expectation vs. experience: Evaluating the usability of code generation tools powered by large language models.
\newblock In {\em Chi conference on human factors in computing systems extended abstracts}, pages 1--7, 2022.

\bibitem{vaswani2017attention}
Ashish Vaswani, Noam Shazeer, Niki Parmar, Jakob Uszkoreit, Llion Jones, Aidan~N Gomez, {\L}ukasz Kaiser, and Illia Polosukhin.
\newblock Attention is all you need.
\newblock {\em Advances in neural information processing systems}, 30, 2017.

\bibitem{Wei2023-CopilotingCopilotsFusing}
Yuxiang Wei, Chunqiu~Steven Xia, and Lingming Zhang.
\newblock Copiloting the {{Copilots}}: {{Fusing Large Language Models}} with {{Completion Engines}} for {{Automated Program Repair}}.
\newblock In {\em Proceedings of the 31st {{ACM Joint European Software Engineering Conference}} and {{Symposium}} on the {{Foundations}} of {{Software Engineering}}}, pages 172--184, November 2023.

\bibitem{weicker1984dhrystone}
Reinhold~P Weicker.
\newblock Dhrystone: a synthetic systems programming benchmark.
\newblock {\em Communications of the ACM}, 27(10):1013--1030, 1984.

\bibitem{Xia2023-ConversationalAutomatedProgram}
Chunqiu~Steven Xia and Lingming Zhang.
\newblock Conversational {{Automated Program Repair}}, January 2023.

\bibitem{xie2006static}
Yichen Xie and Alex Aiken.
\newblock Static detection of security vulnerabilities in scripting languages.
\newblock In {\em USENIX Security Symposium}, volume~15, pages 179--192, 2006.

\bibitem{xu2018manufacturing}
Hui Xu, Yangfan Zhou, Yu~Kang, Fengzhi Tu, and Michael Lyu.
\newblock Manufacturing resilient bi-opaque predicates against symbolic execution.
\newblock In {\em 2018 48th Annual IEEE/IFIP International Conference on Dependable Systems and Networks (DSN)}, pages 666--677. IEEE, 2018.

\bibitem{xu2012power}
Wei Xu, Fangfang Zhang, and Sencun Zhu.
\newblock The power of obfuscation techniques in malicious javascript code: A measurement study.
\newblock In {\em 2012 7th International Conference on Malicious and Unwanted Software}, pages 9--16. IEEE, 2012.

\bibitem{xu2023lmpa}
Xiangzhe Xu, Zhuo Zhang, Shiwei Feng, Yapeng Ye, Zian Su, Nan Jiang, Siyuan Cheng, Lin Tan, and Xiangyu Zhang.
\newblock Lmpa: Improving decompilation by synergy of large language model and program analysis.
\newblock {\em arXiv preprint arXiv:2306.02546}, 2023.

\bibitem{yuan2023evaluating}
Zhiqiang Yuan, Junwei Liu, Qiancheng Zi, Mingwei Liu, Xin Peng, and Yiling Lou.
\newblock Evaluating instruction-tuned large language models on code comprehension and generation.
\newblock {\em arXiv preprint arXiv:2308.01240}, 2023.

\bibitem{zhang2019novel}
Jian Zhang, Xu~Wang, Hongyu Zhang, Hailong Sun, Kaixuan Wang, and Xudong Liu.
\newblock A novel neural source code representation based on abstract syntax tree.
\newblock In {\em 2019 IEEE/ACM 41st International Conference on Software Engineering (ICSE)}, pages 783--794. IEEE, 2019.

\bibitem{zhang2023privee}
Ruoyu Zhang, Ruijie Fang, Chongzhou Fang, Houman Homayoun, and Gozde Goncu~Berk.
\newblock Privee: A wearable for real-time bladder monitoring system.
\newblock In {\em Adjunct Proceedings of the 2023 ACM International Joint Conference on Pervasive and Ubiquitous Computing \& the 2023 ACM International Symposium on Wearable Computing}, pages 291--295, 2023.

\bibitem{zhao2023survey}
Wayne~Xin Zhao, Kun Zhou, Junyi Li, Tianyi Tang, Xiaolei Wang, Yupeng Hou, Yingqian Min, Beichen Zhang, Junjie Zhang, Zican Dong, et~al.
\newblock A survey of large language models.
\newblock {\em arXiv preprint arXiv:2303.18223}, 2023.

\bibitem{zheng2023judging}
Lianmin Zheng, Wei-Lin Chiang, Ying Sheng, Siyuan Zhuang, Zhanghao Wu, Yonghao Zhuang, Zi~Lin, Zhuohan Li, Dacheng Li, Eric Xing, et~al.
\newblock Judging llm-as-a-judge with mt-bench and chatbot arena.
\newblock {\em arXiv preprint arXiv:2306.05685}, 2023.

\bibitem{zhong2017seq2sql}
Victor Zhong, Caiming Xiong, and Richard Socher.
\newblock Seq2sql: Generating structured queries from natural language using reinforcement learning.
\newblock {\em arXiv preprint arXiv:1709.00103}, 2017.

\end{thebibliography}

\end{document}